\documentclass[a4paper,11pt]{article}
\pdfoutput=1 %

\usepackage{jcappub} %

\usepackage[T1]{fontenc} %

\usepackage[utf8]{inputenc}
\usepackage{aas_macros}
\usepackage{amsfonts}
\usepackage{amsmath}
\usepackage{amssymb}
\usepackage[normalem]{ulem}
\usepackage{graphicx}
\usepackage[dvipsnames]{xcolor}
\usepackage{hyperref}
\hypersetup{colorlinks=true,allcolors=teal}

\newcommand{\der}{\ensuremath{{\rm d}}}

\newcommand{\eqn}[1]{equation~\eqref{#1}}
\newcommand{\eqns}[1]{equations~\eqref{#1}}
\newcommand{\eqnref}[1]{equation~\eqref{#1}}

\newcommand{\figref}[1]{Figure~\ref{#1}}
\newcommand{\tabref}[1]{Table~\ref{#1}}
\newcommand{\secref}[1]{Section~\ref{#1}}

\newcommand{\be}{\begin{equation}}
\newcommand{\ee}{\end{equation}}
\newcommand{\Cal}[1]{\ensuremath{\mathcal{#1}}}

\title{\boldmath A self-similar model of galaxy formation and dark halo relaxation}

\author{Premvijay Velmani}
\author{and Aseem Paranjape}
\affiliation{Inter-University Centre for Astronomy \& Astrophysics,\\ Ganeshkhind, Post Bag 4, Pune 411007, India}

\emailAdd{premv@iucaa.in}
\emailAdd{aseem@iucaa.in}

\abstract{
We develop a spherical self-similar model for the formation of a galaxy through gas collapsing in an isolated self-gravitating dark matter halo. As is well known, the self-similarity assumption makes the problem eminently tractable by reducing it to a system of ordinary differential equations. We improve upon the existing literature on self-similar collapse in two ways. First, we include the effects of radiative cooling and the formation of a pseudo-disk at the center of collapse, in a parametrised manner. More importantly, we solve for the evolution of gas and dark matter \emph{simultaneously and self-consistently} using a novel iterative approach. As a result, our model produces shell trajectories of both gas \emph{and} dark matter that qualitatively agree with the results of full hydrodynamical simulations of self-gravitating systems. We discuss the impact of various ingredients such as the accretion rate, gas equation of state, disk radius and cooling rate amplitude on the evolution of the gas shells, although we leave the inclusion of stellar and black hole activity to future work. The self-consistent evolution of gas and dark matter allows us to study the response (or `quasi-adiabatic relaxation') of the dark matter trajectories to the presence of collapsing gas, an effect that has gained increasing importance recently in the context of precision estimates of small-scale statistics like the matter power spectrum. Our default configuration produces a relaxation relation in qualitative agreement with that seen in cosmological hydrodynamical simulations, and further allows us to easily study the impact of the model ingredients mentioned above. As an initial application, we vary one ingredient at a time and find that the accretion rate and gas equation of state have the largest impact on the relaxation relation, while the cooling amplitude plays only a minor role. Our model thus provides a convenient framework to rapidly explore the coupled nonlinear impact of multiple astrophysical processes on the mass and velocity profiles of dark matter in galactic halos, and consequently on observables such as rotation curves and gravitational lensing signals.
}


\begin{document}
\maketitle
\flushbottom

\section{Introduction}
\label{sec:intro}

\noindent
The formation and evolution of galaxies and their associated dark matter halos is one of the primary astrophysical questions being currently addressed both observationally \cite{2020SDSS_Ahumadaetal_16th_data,2017Willetetal_GalxayZoo_HST,2023Harikane_etal_JWST}
and theoretically \cite{2015SomervilleDave,2018WechslerTinker}
Extensive theoretical work has been performed using numerical hydrodynamical simulations of both individual objects \cite{2006Dekel&Birnboim,2014Hopkins_FIRE,2015LauNagaietal,2023FIRE-2_publicrelease} as well as cosmological volumes \cite{2010Schaye_OWLS,2014Genel_Illustris,2015Schaye_EAGLE,2017Kaviraj_HorizonAGN,2018TNG_Pillepich_etal,2019Dave_SIMBA,2021camels_presentation} 
to understand the impact of a variety of astrophysical processes involved in galaxy evolution. From a cosmological viewpoint, these `baryonic' effects are often a nuisance in parameter inference; e.g., the effects of feedback due to an active galactic nucleus (AGN) can be degenerate with the effects of a massive neutrino species or a thermally produced `warm' dark matter candidate at Fourier scales $k\sim1\,h\,{\rm Mpc}^{-1}$ (see, e.g., \cite{2019Chisari_etal_Baryfeedback,2020AricoAnguloetal_baryonifi}).
More generally, the response of the dark matter (DM) to the presence of baryons first must be quantified and then distinguished from the observational effects of (non-standard) primordial physics. While the former problem has a long history \cite{1986Blumenthal,2004Gnesin_etal,2005SellwoodMcGaugh,2006Gustafsson_FS,2010Abadi_NFBS,2010DuffySchaye_etal,2010PedrosaTissera_etal,2010TisseraWhite_etal,2019ArtalePedrosa_etal,2022ForouharMoreno_etal,2023Velmani&Paranjape},
the latter has only recently begun to be studied in detail \cite{2011TeyssierMMDM,2015SchneiderTeyssier,2015Mead_PHJH,2020AricoAnguloetal_baryonifi,2021AricoAnguloetal_baryonifi,2023EuclidCastro_etal}.

The most realistic description of halo response or `relaxation' in the presence of baryons is provided by hydrodynamical simulations that include calibrated effects of star formation, radiative cooling, feedback due to stellar winds and supernovae as well as the triggering of AGN activity in super-massive black holes and its feedback on star formation \cite{2015Schaye_EAGLE,2018TNG_Pillepich_etal}. 
In our earlier work \cite{2023Velmani&Paranjape} we quantified this response in the IllustrisTNG \cite{2018TNG_Pillepich_etal}
and EAGLE \cite{2015Schaye_EAGLE} %
simulation suites by studying the mass profiles of matched pairs of halos in the hydrodynamical and gravity-only runs of the simulation, and described the response in the language of quasi-adiabatic relaxation \cite{2011TeyssierMMDM}. %
In particular, we showed that the original quasi-adiabatic prescription accurately describes the relaxation of halos in simulations provided we include the halo-centric distance in units of the halo radius as an additional parameter. We further provided convenient fitting forms for this parametrised relaxation at $z=0$ and explored its dependence on various halo and galaxy properties such as halo concentration, stellar mass, star formation rate, etc.

Physically, however, the mechanisms driving this halo response are not entirely clear. In principle, this question can also be addressed using hydrodynamical simulations by systematically varying the physical prescriptions defining different processes, particularly the observationally uncertain ones associated with stellar and especially AGN feedback. This, however, is a very computationally expensive exercise requiring multiple runs of expensive simulations \cite{2015Schaye_EAGLE,2021camels_presentation}.
We will report the results of such an exercise using the CAMELS simulation suite \cite{2022camels_data_release1} in a separate publication. In the present work, we investigate whether we can gain some physical insights into the problem of halo relaxation due to baryonic backreaction using a simplified, spherically symmetric toy model of galaxy evolution.

Our key assumption will be that the model is `self-similar', so that spatial variations are related to temporal variations through a scale radius that evolves over time. The self-similar assumption offers a powerful tool to simplify the problem and render it more tractable, by converting nonlinear coupled partial differential equations into ordinary differential equations. The self-similar model was pioneered by Fillmore and Goldreich (1984) \cite{1984FillmoreGoldreich} and Bertschinger (1985) \cite{1985Bertschinger}. While Bertschinger devised spherical self-similar solutions for the collapse of dark matter and shocked gas within the Einstein-de Sitter (EdS) universe, under the assumption of a constant accretion rate, Fillmore and Goldreich explored self-similar solutions for the collapse of dark matter from various initial mass distributions, expanding their purview to include not only spherical but also cylindrical and planar self-similar collapses. Bertschinger (1989) \cite{1989Bertschinger} delved into self-similar cooling flows of gas, focusing on the inner regions of galaxies. Owen et al. (1998) \cite{1998OwenWeinberg_etal} derived a set of cooling functions that guarantee self-similar evolution by ensuring that the cooling time-scale of an object with a characteristic clustering mass remains a constant fraction of the Hubble time. Abadi et al. (2000) \cite{2000Abadi_etal_SelfSimCool} studied the self-similar accretion of gas with radiative cooling in the EdS Universe. Shi (2016a) \cite{2016ShiDMLamCDM} generalised self-similar models to Lambda-cold dark matter ($\Lambda$CDM) models, focusing on the outer profile and, notably, the outermost caustic or `splashback radius' of dark matter collapse which has gained recent popularity \cite{2014DiemerKrastov,2014AdhikariDalalChamberlain,2018Changetal_DES_splashback}.
In a complementary study, Shi (2016b) \cite{2016ShiICM} probed the self-similar accretion of shocked gas, dissecting its behavior with respect to accretion rates and revealing correlations between the shock radius of gas and the dark matter's splashback radius.

The present work draws upon these foundations but incorporates two novel additions: (i) a self-similar cooling of the gas and (ii) the formation of a pseudo-disk of gas along with the self-similar collapse of the dark matter halo. Unlike the work of \cite{1989Bertschinger}, where the cooling scale was set by a cooling radius and was primarily applicable to the innermost regions, in our work we set the cooling scale relative to the turn-around radius. This allows us to keep intact the global self-similarity of the model, while allowing for a range of possible amplitudes for the cooling rate. Our inspiration for including the formation of a pseudo-disk in our spherical approach comes from spherical hydrodynamical simulations that have previously explored the influence of cooling on shock-heated gas, tracing its subsequent accretion onto central structures resembling galaxy disks \cite{2006Dekel&Birnboim}. These simulations have also revealed a `cold mode' of gas accretion that occurs in low mass halos, wherein the gas directly accretes onto an inner disk without shocking. While these results are consistent with those of full hydrodynamical simulations \cite{2005Keres_KWD},
the main relevance to us is the fact that the inner disk radius undergoes an approximately self-similar evolution, following the behavior of the turnaround radius, thus rendering it amenable to our self-similar modeling.

\emph{Importantly, we self-consistently solve for the evolution of the gas and the dark matter simultaneously using a novel iterative technique.} This approach allows us to reliably estimate the relaxation of the dark matter due to baryonic backreaction in this self-similar model. The inclusion of cooling and the formation of a pseudo-disk also brings our self-similar model of galaxy formation substantially closer to the more realistic results of hydrodynamical simulations. We leave the incorporation of the formation of stars and a central black hole, and the associated stellar and AGN feedback, to future work.

The paper is organised as follows. In \secref{sec:methods}, we outline the self-similar model of dark matter halo collapse and collisional gas, followed by a description of our new self-similar galaxy formation model in an EdS universe. In \secref{sec:results}, we discuss the solutions to our self-similar galaxy formation model and its co-evolution with the self-similar collapse of the host dark matter halo, including the impact of variations in the free parameters of our model. We conclude in \secref{sec:conclusion}.

\section{Methods}
\label{sec:methods}
\subsection{Self-similar evolution of dark matter infall}
\label{sec:methods-dm}
We start with an outline of the self-similar spherical collapse model of the self-gravitating dark matter halo in the EdS universe. We describe our model of galaxy-halo co-evolution in the following (sub)sections. The evolution of spherical shells with physical radius $r(t)$ under the gravitational influence of the mass enclosed is given by, 
\begin{align}
\label{eq:dm_traj_phys_accl}
\frac{\der^2 r}{\der t^2} = - \frac{GM(r,t)}{r^2}.
\end{align}
Their trajectories can be solved analytically using Lagrangian coordinate as long as those shells don't cross each other; that is for a shell at radius $r_i$ at some early time $t_i$ with velocity $v_i$, the \eqn{eq:dm_traj_phys_accl} can be integrated until the enclosed mass remains same as the initial mass $M_i$,
\begin{align}
\label{eq:dm_traj_phys_ener}
\left( \frac{\der r}{\der t} \right)^2 - v_i^2 = \frac{2GM_i}{r} - \frac{2GM_i}{r_i}\,,
\end{align}
which simply represents conservation of energy for each shell.
Further by expressing the enclosed mass $M_i$ in terms of the overdensity in the sphere enclosed by the shell above the critical density $\rho_H(t_i)$, we have
\begin{align}
\nonumber
\delta_i \equiv \delta M_i / M_i \implies \frac{2GM_i}{r_i} = \frac{2G \rho_H(t_i)}{r_i} \frac{4}{3}\pi r_i^3 (1+\delta_i) = r_i^2 H_i^2 (1+\delta_i).
\end{align}
where $H_i$ denotes the initial value of the Hubble parameter $H=2/(3t)$ in the EdS universe. Further assuming the shells follow Hubble flow initially
($v_i^2 = r_i^2 H_i^2$), 
the \eqn{eq:dm_traj_phys_ener} gives analytical solution for the trajectory of all the shells \cite{1993paddy_strucformbook},
\begin{gather}
\left( \frac{\der r}{\der t} \right)^2 = \frac{2GM_i}{r} - r_i^2 H_i^2 \delta_i,\\
\label{eq:dm_traj_soln_phys}
r(\theta) = (1 - \cos \theta) G M_i/r_i^2 H_i^2 \delta_i \qquad t(\theta) = (\theta - \sin \theta) G M_i/r_i^3 H_i^3 \delta_i^{3/2}.
\end{gather}
Each shell expands initially but then turns around at a radius of $r_{\smallfrown}= 2G M_i/r_i^2 H_i^2 \delta_i$ at time $t_{\smallfrown} = \pi G M_i/r_i^3 H_i^3 \delta_i^{3/2}$ that is specific to that shell. 
However, as a dark matter shell falls into the halo after this turnaround,
its trajectory starts to deviate from this solution when crossing other collisionless dark matter shells that are expanding outwards from the centre. 
Moreover, these shells turn around once again and start accreting as a secondary infall. This leads to the formation of caustics such as the splashback radius. 

Even in such regions, self-similar solutions are obtained assuming scale-free initial perturbation with a power law form of overdensity distribution $\delta_i \propto M_i^{-\epsilon}$ \cite{1984FillmoreGoldreich}. For sufficiently larger scales, with small perturbation $\delta_i \ll 1$, we can write $M_i \propto r_i^3$ and hence 
$r_{\smallfrown} \propto M_i^{1/3+\epsilon}$ \& $t_{\smallfrown} \propto M_i^{3\epsilon/2}$. At any given time $t$, the shell that undergoes turnaround, encloses a mass of $M_{\smallfrown} \propto t^{2/3\epsilon} \propto a^{s}$, where $s=1/\epsilon$ is the mass accretion rate of the halo; hence this parameter $s$ relates to both the nature of initial perturbation and the accretion rate. It follows that the (turnaround) radius of such a shell undergoing turnaround at any time $t$, has a simple power law evolution $r_{\smallfrown} \propto t^{\delta}$ with $\delta \equiv 2(1+s/3)/3$. We denote this as $r_{\smallfrown}(t)$ to distinguish it from the $r_{\smallfrown}$ which is a constant attached to each shell. This $r_{\smallfrown}(t)$ is then used to set the self-similar scale of the system; then for every shell whose trajectory is given by the \eqn{eq:dm_traj_phys_accl}, the evolution of its radius characterized by $\lambda \equiv r/r_{\smallfrown}(t)$ can be described as follows,
\begin{align}
\label{eq:traj_DM_selfsim}
\frac{\der^2 \lambda}{\der \xi^2} + (2 \delta - 1) \frac{\der \lambda}{\der \xi} + \delta ( \delta -1) \lambda = - \frac{2}{9} \frac{\mathcal{M}(\lambda)}{\lambda^2 }
\end{align}
where $\xi \equiv \ln \tau $ with $ \tau \equiv t/t_{\smallfrown}$ and $t_{\smallfrown}$ being the turnaround time of that shell \cite{1985Bertschinger,2016ShiICM}. 
Here the dimensionless self-similar mass profile $\mathcal{M}(\lambda)$ is defined such that,
\begin{align}
M(r=\lambda r_{\smallfrown}(t),t) = \rho_H \left( \frac{4}{3} \pi (r_{\smallfrown}(t))^3 \right) \mathcal{M}(\lambda) %
\end{align}
which is given by integrating the infinitesimal mass carried in corresponding shells as 
\begin{align}
\label{eq:mass_integ_DM}
\mathcal{M}(\lambda) = \int_{0}^{M_{\rm{ta}}} \frac{\der M_i}{M_{\rm{ta}}} H[\lambda - \lambda'(\xi')] = \int_{1}^{\infty} d \xi' e^{2s\xi'/3} H[\lambda - \lambda'(\xi')],
\end{align}
where $H(x)$ is the Heaviside step function.

For an initial guess of the mass profile $\mathcal{M}(\lambda)$, the differential \eqn{eq:traj_DM_selfsim} is solved for the shell trajectory, which is then integrated using \eqn{eq:mass_integ_DM} to obtain a corresponding mass profile. The trajectory in this updated mass profile is then solved to update the trajectory; following this iteratively we obtain converged solution for the self-similar collapse of dark matter.

\subsection{Self-similar evolution of galaxy formation}
For spherical shells of collisional gas accreting due to self-gravity, the \eqn{eq:gas_traj_phys_accl} gets an additional pressure term,
\begin{align}
\label{eq:gas_traj_phys_accl}
\frac{\der^2 r}{\der t^2} = - \frac{GM(r,t)}{r^2} -\frac{1}{\rho(r,t) }\frac{\partial p(r,t)}{\partial r}.
\end{align}
In the absence of shell-crossing for collision gas, the evolution of velocity profiles along with the density, pressure and mass profiles is given by the spherical hydrodynamical equations \cite{1985Bertschinger} as
\begin{align}
\label{eq:sph_hydro_cont}
\frac{{\rm d}\rho }{{\rm d}t} &= -\frac{\rho }{r^2}\frac{\mathrm{\partial} }{\mathrm{\partial} r} \left( r^2v \right) , \\
\label{eq:sph_hydro_accl}
\frac{{\rm d}v}{{\rm d}t} &= - \frac{GM(r,t)}{r^2} -\frac{1}{\rho }\frac{\partial p}{\mathrm{\partial} r}, \\
\label{eq:sph_hydro_mass_cons}
\frac{\mathrm{\partial} M}{\mathrm{\partial} r} &= 4\pi r^2 \rho.
\end{align}
In addition to these equations, we have the energy conservation equation for adiabatic accretion of gas, 
\begin{align}
\label{eq:sph_hydro_adiab}
\frac{{\rm d}}{{\rm d}t}\left( p \rho ^{-\gamma } \right) = 0.
\end{align}
Gas shells have negligible pressure as they initially expand with the Hubble flow and hence they follow the same trajectory initially as described by \eqn{eq:dm_traj_soln_phys} turn around similar to the collisionless dark matter. Hence, self-similar radius $\lambda$ is defined as described in the \secref{sec:methods-dm} for dark matter; and the dimensionless self-similar profiles of velocity $V(\lambda)$, density $D(\lambda)$, mass $\mathcal{M}(\lambda)$ and pressure $P(\lambda)$ are then defined such that,
\begin{align*}
\rho &= D(\lambda ) \rho_H = \frac{D(\lambda )}{6 \pi G t^2}, \qquad
p = P(\lambda ) \rho_H \left[ \frac{r_{\smallfrown}(t)}{t} \right]^2,\\
v &= \frac{r_{\smallfrown}(t)}{t} V(\lambda ), \quad M = \frac{4\pi \rho _{\rm H}}{3} (r_{\smallfrown}(t))^3\,\mathcal{M}(\lambda)\quad  \text{with} \quad r_{\smallfrown}(t) \propto t^{\delta};
\end{align*}
The hydrodynamics of the self-similar collapse of adiabatic gas is then given by ordinary differential equations over these  profiles,
\begin{align}
\label{eq:selfsim-sph_hydro_cont}
(V-\delta \lambda) D' + DV' + \frac{2 D V}{\lambda} - 2D &= 0\\
\label{eq:selfsim-sph_hydro_accl}
(V-\delta \lambda) V' + (\delta -1 ) V &= - \frac{2}{9} \frac{\mathcal{M}(\lambda)}{\lambda^2 } - \frac{P'}{D}\\
\label{eq:selfsim-sph_hydro_adiab}
(V-\delta \lambda) \left( \frac{P'}{P} - \gamma \frac{D'}{D} \right) &= -2 (\gamma - 1) - 2 (\delta -1)\\
\label{eq:selfsim-sph_hydro_mass_cons}
\mathcal{M}' &= 3 \lambda^2 D
\end{align}
where a prime denotes the first-order derivative of the corresponding self-similar quantity with respect to the self-similar radius $\lambda$.

\subsubsection{Shock formation}%
As the gas shells turn around and free fall to the centre, they are assumed to get shocked at a radius proportional to the turnaround radius of the shell turning around at that time. Such a shock maintains the self-similarity by 
propagating outwards as the halo grows by accreting mass.
In the self-similar gas profiles, this shock remains at a fixed radius $(\lambda_{\rm s})$, outside which the gas profiles are evolved pressurelessly starting from the turn-around radius;
and inside $\lambda_{\rm s}$ the gas is evolved with pressure support generated by the shock.
For such a self-similarly propagating shock, the jump conditions \cite{1985Bertschinger} expressed in terms of the dimensionless self-similar quantities are 
\begin{eqnarray}
\label{eq:shock_jump_selfsim}
V_2 &=& \frac{\gamma -1}{\gamma +1} \left[ V_1-\delta \lambda _{\rm s} \right] + \delta \lambda _{\rm s} \,,\nonumber\\
D_2 &=& \frac{\gamma +1}{\gamma -1} D_1 \,,\nonumber\\
P_2 &=& \frac{2}{\gamma +1} D_1 \left[ V_1-\delta \lambda _{\rm s} \right]^2 \,,\nonumber\\
M_2 &=& M_1 \,
\end{eqnarray}
where the subscripts 1 and 2 denote the value of the corresponding quantity outside and inside the shock, respectively, and $\gamma$ is the adiabatic equation of state of the gas. %
For a given shock radius, these relations give the post-shock quantities from the pre-shock quantities obtained by pressureless evolution. 
The complete solution for self-similar collapse of adiabatic gas is then obtained by solving \eqns{eq:selfsim-sph_hydro_cont}, \eqref{eq:selfsim-sph_hydro_accl}, \eqref{eq:selfsim-sph_hydro_adiab}, and \eqref{eq:selfsim-sph_hydro_mass_cons} with these boundary conditions inside the shock.
Usually, this shock radius $(\lambda_{\rm s})$ is fine-tuned such that gas shells accrete all the way to the center but sufficiently slow to halt at $\lambda=0$ preventing the formation of a black hole \cite{1985Bertschinger}. In our work, we break this adiabaticity by allowing the gas to cool and further form a galaxy-like structure as described in the following sections.

\subsubsection{Cooling of shock heated gas}%
In this section, we outline the incorporation of gas cooling into our self-similar galaxy formation model. The hot gas within the shock is allowed to cool radiatively with a luminosity \Cal{L} per unit volume. 
For a small patch of gas occupying volume \Cal{V}, energy conservation dictates $- \Cal{L}\Cal{V}~dt = dU + p~d\Cal{V}$; where the thermal energy for an ideal gas is expressed as $U = p \Cal{V}/(\gamma-1)$, yielding the following expression:
\begin{align*}
- \Cal{L}\Cal{V}~\der t &= \der \left( \frac{p \Cal{V}}{\gamma-1} \right) + p~d\Cal{V} = \frac{\Cal{V}~\der p}{\gamma-1} + \frac{\gamma p~d\Cal{V}}{\gamma -1}\\
\implies - \frac{\gamma -1}{p} \mathcal{L} ~\der t &= \frac {\der p}{p} + \gamma \frac {\der \Cal{V}}{\Cal{V}} = \der \ln \left( p \Cal{V}^{\gamma} \right).
\end{align*}
Expressing this relationship in terms of the local density, we arrive at the key equation,
\begin{align}
\label{eq:sph_hydro_cooling}
\frac{\der }{\der t} \ln \left( p \rho^{-\gamma} \right) &= - \frac{\gamma -1}{p}\mathcal{L}.
\end{align}
Here, the luminosity function is chosen to have a power-law form to produce self-similar cooling \cite{1989Bertschinger},
\begin{align}
\label{eq:cooling_luminosity}
\mathcal{L} = \rho^2 \Lambda_0 \left( \frac{p}{\rho} \right)^{\nu}.
\end{align}
Inserting this in the \eqn{eq:sph_hydro_cooling} gives the cooling equation which yields the following equation for the self-similar dimensionless quantities $(V,D,P)$:
\begin{align}
\label{eq:selfsim-sph_hydro_cooling}
(V-\delta \lambda) \left( \frac{P'}{P} - \gamma \frac{D'}{D} \right) &= (-\Bar{\Lambda}_0 D^{(2-\nu)} P^{\nu-1} - 2) (\gamma -1) - 2(\delta-1)
\end{align}
where 
\begin{align}
\Bar{\Lambda}_0 &\equiv \Lambda_0 \rho_H \left( \frac{r_{\smallfrown}(t)}{t} \right)^{2(\nu-1)} t.
\end{align}
In order to produce a self-similar cooling gas solution scaled by the turnaround radius, we have to assume that $\Bar{\Lambda}_0$ is constant. Since we have, $\rho_H \propto t^{-2} ~ \& ~
r_{\smallfrown}(t) \propto t^{\delta}$, this implies that for self-similar cooling we need
\begin{align}
\Bar{\Lambda}_0  \propto \Lambda_0 t^{-2} t^{2(\delta-1)(\nu-1)} t = \text{constant}
\end{align}
Note that here, the $\Lambda_0$ can have implicit time dependence. For example when $s=1.5$, we have $\delta=1$, and hence we require that $\Lambda_0 \propto t$; such a time dependence on the cooling function amplitude could come from changing metallicity, etc. For other values of accretion rate self-similar solutions are possible even with a constant $\Lambda_0$ by choosing the power-law parameter accordingly. Restricting to self-similar solutions, we still have two parameters $\Bar{\Lambda}_0$ and $\nu$ that allow this model to explore different cooling behaviors. For any given choice of these cooling parameters, the gas profiles can be obtained using \eqns{eq:selfsim-sph_hydro_cont}, \eqref{eq:selfsim-sph_hydro_accl}, \eqref{eq:selfsim-sph_hydro_cooling}, and \eqref{eq:selfsim-sph_hydro_mass_cons}.

\subsubsection{Galaxy pseudo-disk formation}%
To mitigate the formation of massive black holes resulting from the accretion of cooled shells toward the center, we introduce a corrective artificial viscosity-like force term in this study. We assume this force is proportional to $vr^{-\nu_d}$, where $\nu_d>3$, so that it becomes dominant in the inner region ($r \rightarrow 0$) when shells exhibit high infall velocities. As the shells decelerate, the force diminishes, approaching zero as the velocity ($v$) tends to zero. In the context of self-similar quantities, the modified acceleration equation, accounting for this additional force term, is represented as follows:
\begin{align}
\label{eq:selfsim-sph_hydro_accl_disk_gen}
(V-\delta \lambda) V' + (\delta -1 ) V &= - \frac{2}{9} \frac{\mathcal{M}(\lambda)}{\lambda^2 } - \frac{P'}{D} + k_d V \lambda^{-\nu_d}.
\end{align}
We choose a substantial value for $\nu_d$, of $\nu_d=10$, so that this term dominates only in the inner region. Through empirical investigation, we determine that this resembles spherically averaged profiles in the case of
accretion onto a galaxy disk-like structure. The size of this structure can be manipulated by adjusting the amplitude of the viscosity term. We find that a self-similar radius of $\lambda_{\rm d} \equiv (k_d/10)^{10}$, reasonably characterizes the size of this pseudo-disk. We call this structure a pseudo-disk as it effectively captures the effect of a disk in this one-dimensional spherically symmetric model.  In \eqn{eq:selfsim-sph_hydro_accl_disk_gen}, expressing the amplitude of the force term in relation to the size ($\lambda_{\rm d}$) of the resulting disk-like structure, we obtain:
\begin{align}
\label{eq:selfsim-sph_hydro_accl_disk}
(V-\delta \lambda) V' + (\delta -1 ) V &= - \frac{2}{9} \frac{\mathcal{M}(\lambda)}{\lambda^2 } - \frac{P'}{D} + 10V(\lambda/\lambda_{\rm d})^{-10}.
\end{align}

\noindent
Gas profiles in this model of cooling and accretion of shock-heated gas onto a galaxy disk-like structure are obtained by solving \eqns{eq:selfsim-sph_hydro_cont}, \eqref{eq:selfsim-sph_hydro_accl_disk}, \eqref{eq:selfsim-sph_hydro_cooling}, and \eqref{eq:selfsim-sph_hydro_mass_cons} inside the shock; these equations can be expressed as,

\begin{align}
\label{eq:selfsim-sph_hydro-mat-gaso}
\begin{split}
\frac{\der \ln}{\der \ln \lambda}
\begin{bmatrix}
 -\bar{V}\\
 D\\
 P
\end{bmatrix} &= \frac{1}{\bar{V}^2} \frac{1}{\gamma \mathcal{T}-1}
\begin{bmatrix}
-\gamma \mathcal{T} & 1 & -\mathcal{T}\\
1 & -1 & \mathcal{T}\\
\gamma & -\gamma & 1
\end{bmatrix} \begin{bmatrix}
 2 \bar{V} (V-\lambda) \\
\frac{2}{9} \mathcal{M} \lambda^{-1} + (\delta -1 ) V \lambda - 10V(\lambda/\lambda_{\rm d})^{-10} \\
 \bar{V} \lambda [(2-\Bar{\Lambda}_0 D^{(2-\nu)} P^{\nu-1})(\gamma - 1) + 2(\delta -1)]
\end{bmatrix} \\
 &\qquad \qquad \qquad  \qquad  -\begin{bmatrix}
\delta \lambda/ \bar{V}\\
0\\
0\\
\end{bmatrix}\\
\frac{\der \ln}{\der \ln \lambda} \mathcal{M} &= 3 \tilde{D} \lambda^{3} /  \mathcal{M}
\end{split}
\end{align}
where thermal to kinetic ratio $\mathcal{T} \equiv P / D\bar{V}^2$ and 
\be
\bar{V} \equiv \frac{\der \lambda}{\der \xi} = V - \delta \lambda\,.
\label{eq:Vbar-def}
\ee

\subsection{Interplay of galaxy and dark matter halo}
\label{sec:methods-interplay}
When we include both a pressureless dark matter and gas, their evolution gets coupled only by gravity which is given by the total enclosed mass $\mathcal{M}_g(\lambda)+\mathcal{M}_d(\lambda)$. The self-similar evolution of gas profiles in the presence of dark matter profile is
\begin{align}
\label{eq:selfsim-sph_hydro-mat-gasdm}
\begin{split}
\frac{\der \ln}{\der \ln \lambda}
\begin{bmatrix}
 -\bar{V}\\
 D\\
 P
\end{bmatrix} &= \frac{1}{\bar{V}^2} \frac{1}{\gamma \mathcal{T}-1}
\begin{bmatrix}
-\gamma \mathcal{T} & 1 & -\mathcal{T}\\
1 & -1 & \mathcal{T}\\
\gamma & -\gamma & 1
\end{bmatrix}\\
& \times \begin{bmatrix}
 2 \bar{V} (V-\lambda) \\
\frac{2}{9} (\mathcal{M}_g(\lambda)+\mathcal{M}_d(\lambda)) \lambda^{-1} + (\delta -1 ) V \lambda - 10V(\lambda/\lambda_{\rm d})^{-10} \\
 \bar{V} \lambda [(2-\Bar{\Lambda}_0 D^{(2-\nu)} P^{\nu-1})(\gamma - 1) + 2(\delta -1)]
\end{bmatrix} -
\begin{bmatrix}
\delta \lambda/ \bar{V}\\
0\\
0\\
\end{bmatrix}\\
\frac{\der \ln}{\der \ln \lambda} \mathcal{M}_g &= 3 \tilde{D} \lambda^{3} /  \mathcal{M}_g \qquad \qquad
\end{split}
\end{align}
Similarly the evolution dark matter shells is now influenced by the total enclosed mass which includes the gas mass as 
\begin{align}
\label{eq:gas-dm-dm-traj}
\frac{\der^2 \lambda}{\der \xi^2} + (2 \delta - 1) \frac{\der \lambda}{\der  \xi} + \delta ( \delta -1) \lambda = - \frac{2}{9} \frac{(\mathcal{M}_g(\lambda)+\mathcal{M}_d(\lambda))}{\lambda^2 }
\end{align}
We solve this coupled evolution by an iterative procedure. Starting with dark matter only solution for the dark matter mass profile, we solve the hydrodynamical equations to obtain the gas profiles. Then we add the gas mass profile to \eqn{eq:gas-dm-dm-traj} and recompute the dark matter shell trajectory and hence the correponding mass profile by integrating using \eqn{eq:mass_integ_DM}. The updated dark matter mass is then used to solve for the gas profile and the iteration is continued. As we will show below, this iterative procedure converges to self-consistent profiles for dark matter and gas. By comparing the solutions of the dark matter mass profile obtained in such systems with the mass profile obtained in the absence of galaxy, we characterize the response of the halo to galaxy formation.

\subsubsection{The relaxation relation}
\label{sec:methods-relx-reln}
The influence of galaxy formation on the dark halo is often interpreted as a process of (quasi-)adiabatic relaxation, where the orbits of dark matter particles respond to the condensation of baryons \citep[][]{1986Blumenthal,2010Abadi_NFBS,2011TeyssierMMDM,2023Velmani&Paranjape}. Assuming that the dark matter halo maintains a spherical shape and avoids shell crossing as baryons condense toward the center, the extent of adiabatic relaxation for a specific dark matter shell is determined by the change in baryonic mass within that shell. Consider a shell at radius $r_i$ that encompasses a mass of \emph{dark matter} denoted as $M_i^d(r_i)$ within the unrelaxed halo. Following the relaxation process, the shell's radius changes to $r_f$. According to the definition, the dark matter mass, $M_f^d(r_f)$, enclosed at $r_f$ in the relaxed halo is simply given by
\begin{equation}
M_f^d(r_f) = M_i^d(r_i).
\label{eq:DMmass}
\end{equation}
On the other hand, the \emph{total} mass, denoted as $M_i(r_i)$, enclosed at $r_i$ in the unrelaxed halo, does not necessarily equal the total mass, $M_f(r_f)$, enclosed at $r_f$ in the relaxed halo. If angular momentum were to be conserved and the orbits of dark matter particles remained circular, then the degree of relaxation of the shell is entirely determined by the change in this total mass within the shell, as described by \citep[][]{1986Blumenthal}:
\begin{align}
r_i \,M_i(r_i) = r_f \,M_f(r_f) %
\implies 
\frac{r_f}{r_i} = \frac{M_i(r_i)}{M_f(r_f)}.
\label{eq:AR}
\end{align}
Building upon this simplified scenario, quasi-adiabatic relaxation models delve into the relationship between the relaxation ratio, denoted as $r_f/r_i$, and the mass ratio, expressed as $M_i/M_f$, given by \citep[]{2010Abadi_NFBS,2011TeyssierMMDM,2021ParanjapeTirthSheth,2023Velmani&Paranjape} %
\begin{align}
\frac{r_f}{r_i} &= 1 + \chi \left( \frac{M_i(r_i)}{M_f(r_f)} \right) .
\label{eq:qAR}
\end{align}
In a prior study \cite{2023Velmani&Paranjape}, we have demonstrated that this relationship exhibits a local linear behaviour in a wide range of haloes within cosmological hydrodynamical simulations incorporating state-of-the-art baryonic astrophysics. In the present work, we aim to characterize this response by comparing the mass profiles between the self-similar collapse of a self-gravitating dark matter halo and the self-similar collapse of a dark matter halo co-evolving with a self-similar galaxy-like structure through the relation between relaxation ratio 
$r_f/r_i=\lambda_f/\lambda_i$ and mass ratio, $M_i/M_f=\mathcal{M}_i(\lambda_i)/\mathcal{M}_f(\lambda_f)$.

\paragraph{Parameters in the model:}
Parameters in the self-similar model of cooling and accretion of shocked gas onto a galaxy disk-like structure co-evolving with the self-similar collapse of host dark matter halo include the following:
the accretion rate parameter $s$, equation of state parameter of gas $\gamma$, cooling rate amplitude $\Bar{\Lambda}_0$, cooling function slope $\nu$, radius of galaxy disk-like structure $\lambda_{\rm d}$, shock radius $\lambda_{\rm s}$ (see \tabref{tab:parameters-descr}).

\begin{table}[htbp]
\centering
\begin{tabular}{c|c|c|c}%
\hline
Parameter & Description & reference value & variation \\
\hline &&&\\
$s$ & mass accretion rate & 1 & 0.5 to 3 \\
$\gamma$ & gas equation of state & 5/3 & 1.33 to 2  \\
$\lambda_{\rm s}$ & shock radius & 0.9 $\lambda_{\rm sp}$ & 0.5 - 1.1 $\lambda_{\rm sp}$ \\
$f_b$ & baryon fraction within turnaround & 0.1568 & $-$\\
$\Bar{\Lambda}_0$ & dimensionless cooling rate & $0.03$ & $10^{-3}$ to $0.3$ \\
$\nu$ & cooling function slope parameter & 1/2 & \\
$\lambda_{\rm d}$ & radius of disk-like model galaxy & 5\% $\lambda_{\rm s}$ & 2 - 25 \% $\lambda_{\rm s}$  \\
\hline 
\end{tabular}
\caption{Parameters in the self-similar model of cooling and accretion of shock gas onto a galaxy disk-like structure co-evolving with the self-similar collapse of host dark matter halo. Here $\lambda_{\rm sp}$ is the corresponding splashback radius of dark matter.}
\label{tab:parameters-descr}
\end{table}

\section{Results}
\label{sec:results}

\subsection{Gas accretion onto galaxy}
\label{sec:results-gaso}
\begin{figure}[htbp]
\centering
\includegraphics[width=0.665\linewidth]{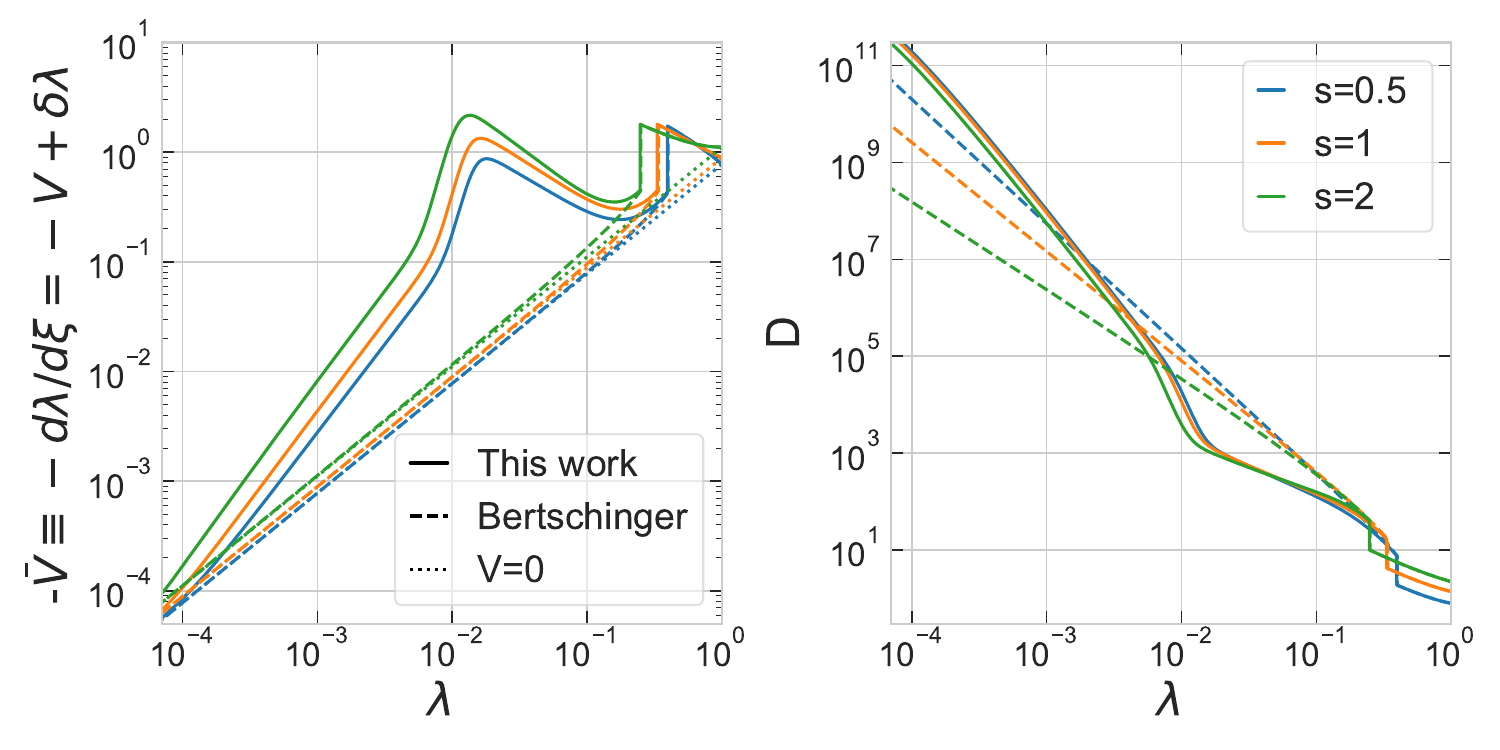}
\includegraphics[width=0.325\linewidth]{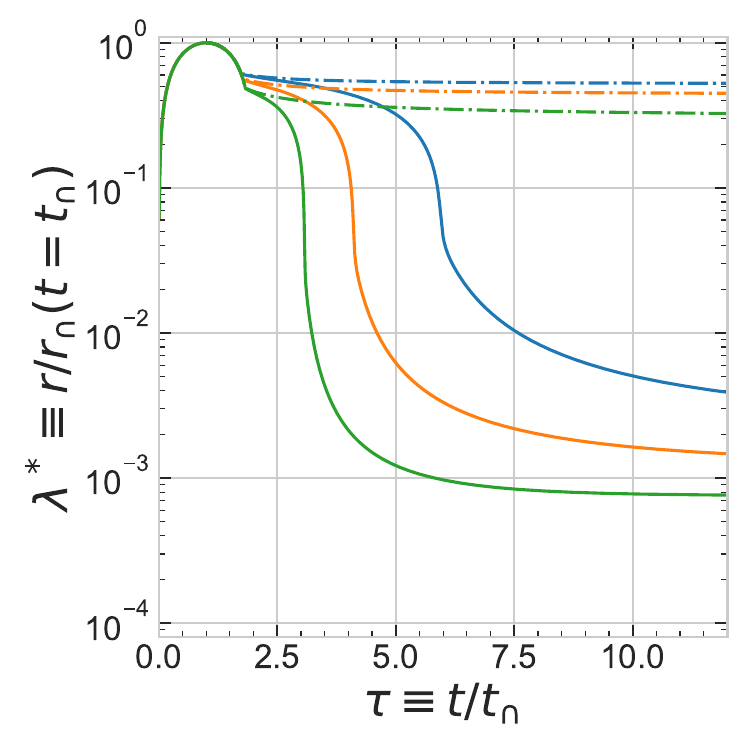}
\caption{%
Self-similar profiles of infall velocity characterized by $-\bar{V}$ (equation~\ref{eq:Vbar-def}) and density are shown in the left and middle panels respectively, while the corresponding trajectories of gas shells are shown in the $(\lambda^*,\tau)$-plane in the right panel. Here the solid curves represent the self-similar evolution in our model for gas with equation of state $\gamma=5/3$, shock heated at 90\% of the corresponding dark matter splashback radius followed by cooling at a rate given by $\bar{\Lambda}_0=3 \times 10^{-2}$, accreting onto a galaxy disk-like structure at $5\%$ of the shock radius under self-gravity in the absence of dark matter background. For comparison the solution to gas profiles in the absence of cooling and galaxy formation is shown by dashed curves \cite{1985Bertschinger}. 
In both cases, the shells come to a halt physically as the velocity reaches the dotted line corresponding to $V=0$.}
\label{fig:gaso-vary-s}
\end{figure}

In this section, we delve into the behaviour of self-similar gas accretion in our galaxy formation model, excluding the presence of dark matter. The effect of co-evolving with a self-similar dark matter halo is explored in \secref{sec:results-gas-dm-coevolve}. In this particular scenario, self-gravitating gas undergoes pressureless accretion outside a self-similar shock, which generates pressure as described by \eqnref{eq:shock_jump_selfsim}. The hot gas supported by this pressure inside the shock subsequently cools and accretes onto a galaxy disk-like structure, following the dynamics governed by \eqnref{eq:selfsim-sph_hydro-mat-gaso}. To illustrate the overall behaviour, we will focus on a specific case using the reference parameter values outlined in Table \ref{tab:parameters-descr}. The self-similar solution in this context is depicted by the orange solid curves in \figref{fig:gaso-vary-s}, where the left panel illustrates the inward velocity profile characterized by $-\bar{V}$ (equation~\ref{eq:Vbar-def}) and the middle panel represents the density profile.

For comparison, the self-similar accretion of pressure-supported adiabatic gas, as studied in previous works \cite{1984FillmoreGoldreich,1985Bertschinger,2016ShiICM}, is represented by dashed orange curves. Despite both cases experiencing shock around the same radius, our implementation of cooling results in shells accelerating to very high velocities as they approach the center. However, the additional viscosity-like term causes a sudden slowdown as they reach a radius of $\lambda_{\rm d}=5\% \lambda_{\rm s} \sim 10^{-2}$. This is followed by a gradual decrease in velocity reaching the dotted line indicating $-\bar{V}=\delta \lambda$ (shown in the left panel of \figref{fig:gaso-vary-s}). 
Throughout this process, the cooling and the formation of the disk-like structure in our model lead to a significant transfer of gas mass within the shock towards the inner disk, as depicted by the density profile in the middle panel of \figref{fig:gaso-vary-s}.

\begin{figure}[htbp]
\centering
\includegraphics[width=0.6\linewidth]{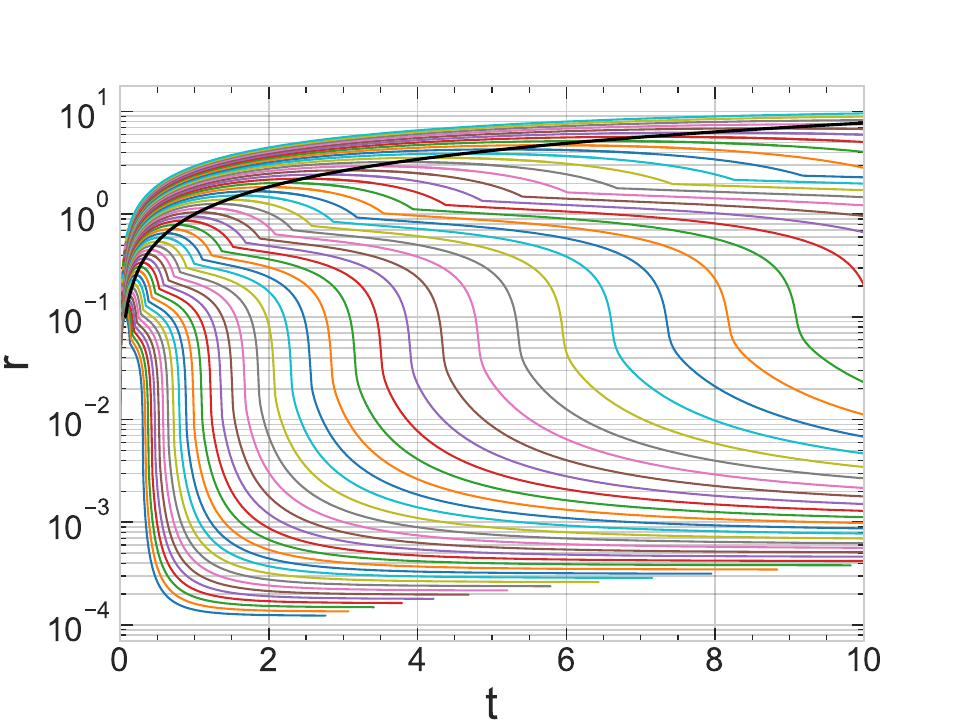}
\caption[]{Physical trajectory of gas shells accreting onto a disk-like structure in our model with reference parameter values as mentioned in \figref{fig:gaso-vary-s}. The solid black curve depicts the evolution of turnaround radius $r_{\smallfrown}(t)$ over time $t$. At any given time, there is a specific shell that undergoes turnaround, while there is another shell that gets shocked at that time. Notice that at all times the ratio of the radius of the shell that gets shocked and that of the shell that is undergoing turnaround is constant with time and a parameter in the model. Similarly, the radius at which the disk-like structure forms also evolves self-similarly with the turnaround radius. Notably, this behavior closely mirrors the spherical hydrodynamical simulations of individual haloes \cite{2006Dekel&Birnboim}.}
\label{fig:gas-traj-phys}
\end{figure}

In the self-similar context, individual gas shells follow unique trajectories in physical coordinates $(r,t)$ (shown in \figref{fig:gas-traj-phys}), but they converge onto a single curve in scaled self-similar coordinates $(\lambda,\xi)$. This shared trajectory, denoted as $\lambda(\xi)$, is derived by integrating the velocity profile $\bar{V} \equiv \frac{\der \lambda}{\der \xi}$ starting with $\lambda(\xi=0)=1$ at the turnaround. However, in these coordinates, the radius ($\lambda$) of a gas shell continues to decrease even after reaching zero physical velocity (when $V=0$) since $\frac{\der \lambda}{\der \xi} = -\delta \lambda \neq 0$. This is due to the fact that, as the halo continues to accrete mass and grow, the turnaround radius $r_{\smallfrown}(t)$ also increases over time, which defines $\lambda$. As an alternative method to characterize the self-similar trajectory, $\lambda^*$ is defined, where the radius of a given shell is scaled at all times by its own (fixed) turnaround radius, rather than the current turnaround radius that evolves with time. Similar to the $(\lambda,\xi)$ coordinates, the trajectory of all shells follows a single curve in the $(\lambda^*,\tau)$ coordinates, expressed as $\lambda^*(\tau) = \lambda(\xi=\ln \tau) \tau^{\delta}$. This is shown in the right panel of \figref{fig:gaso-vary-s} for our model with reference parameters by the orange solid curve. Once again comparing to the simple pressure-supported adiabatic evolution, the gas shells in our model undergo cooling, enabling them to accrete onto almost three orders of magnitude more inner regions, thereby forming the disk-like structure.

The self-similar solution in this galaxy formation model, with higher and lower accretion rates, is also depicted in \figref{fig:gaso-vary-s} by the green ($s=2$) and blue ($s=0.5$) solid curves, respectively. While keeping the remaining model parameters fixed at the reference values, we observe that the gas shells reach a deeper physical radius with faster accretion rates (shown in the left panel). Although the physical size of the disk-like structure in our model (evolving self-similarly with the turn-around radius) grow faster with higher accretion rates, the gas shells also reach this pseudo-disk at an earlier time when it was smaller. This earlier arrival allows them to reach more inner regions (illustrated in the left panel of \figref{fig:gaso-vary-s}). In the middle panel of \figref{fig:gaso-vary-s}, we find that the density profiles are quite similar at different accretion rates in the presence of cooling and galaxy formation in our model. This contrasts with the pressure-supported adiabatic gas, where the inner density profile strongly depends on the accretion rate.

\subsubsection{Shock location}
Previous works ignoring cooling and disk formation have used the inner boundary condition to determine the shock radius $\lambda_{\rm s}$ of the gas \cite{1985Bertschinger}. In such models, the shock is placed at a fine-tuned radius in order to absorb most but not all of the kinetic energy of the infalling gas into thermal energy so that it still reaches the center without forming a black hole. Even a crude approximation of setting post-shock velocity $V_2$ to be zero in the first shock jump \eqn{eq:shock_jump_selfsim}, we expect that such a shock needs to be placed at more inner radius for faster accretion rates and softer gas equation states. Such a trend is in fact seen for the fine-tuned value of the shock radius in the self-similar collapse of adiabatic gas in the absence of any galaxy formation; in particular for gas with the equation of state $\gamma=5/3$, this shock radius is close to the splashback radius of the self-similar collapse of dark matter which is also known to become smaller with faster accretion rate \cite{2016ShiICM}. It has been found that even in full hydrodynamical simulation this shock radius is closely tied to the dark matter splashback radius \cite{2015LauNagaietal}.

\begin{figure}[htbp]
\centering\includegraphics[width=0.665\linewidth,trim={0 0cm 0 0},clip]{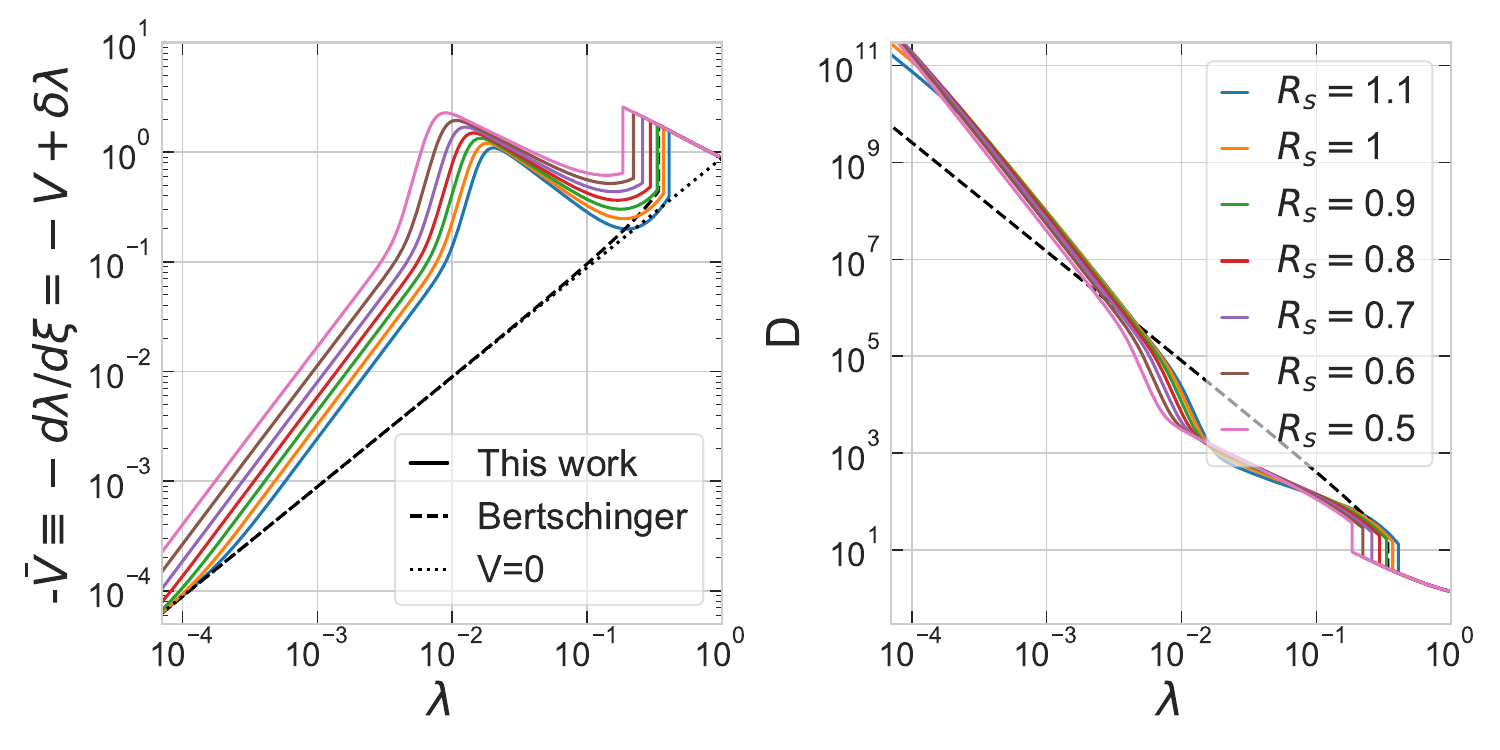}
\includegraphics[width=0.325\linewidth,trim={0 0cm 0 0},clip]{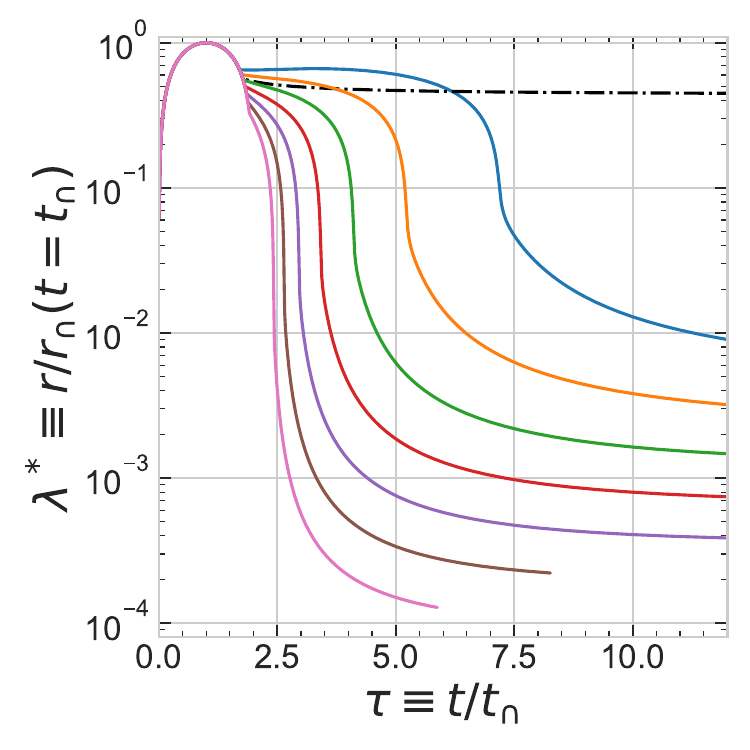}
\caption{Self-similar gas solutions are depicted, akin to those in \figref{fig:gaso-vary-s}. Here, we vary the position of the shock as a fraction $R_{\rm s}$ of the corresponding dark matter splashback radius, while maintaining a fixed accretion rate of $s=1$.}
\label{fig:gaso-vary-lamshsp}
\end{figure}

In our model, on the contrary, we can set the shock radius arbitrarily and still form a galaxy disk-like structure. This is demonstrated in the \figref{fig:gaso-vary-lamshsp}, where the self-similar solution with different choices on the location of the shock denoted by $R_{\rm s} = \lambda_{\rm s}/ \lambda_{\rm sp}$ is shown. When the shock happens at an inner radius, it will not sufficiently slow down the gas shells; however it just accretes quickly onto the galaxy disk-like structure rather than forming a blackhole. This can be seen in the right panel of \figref{fig:gaso-vary-lamshsp} with shells reaching much smaller radii but at an earlier time $\tau$ for smaller values of $R_{\rm s}$. On the other hand, when the gas shells get shocked at a larger radius, even if the shells come to a halt after the shock, cooling in our models allows the gas to continue accreting on the galaxy. As an extreme case when $R_{\rm s}=1.1$, the shells physically start expanding after crossing the shock. This is illustrated in the left panel of \figref{fig:gaso-vary-lamshsp}, indicated by an inward velocity ($-\bar{V}$) going below the dotted line (corresponding to $V=0$) briefly before cooling brings it back allowing the shells to accrete further.

\begin{figure}[htbp]
\centering\includegraphics[width=0.665\linewidth]{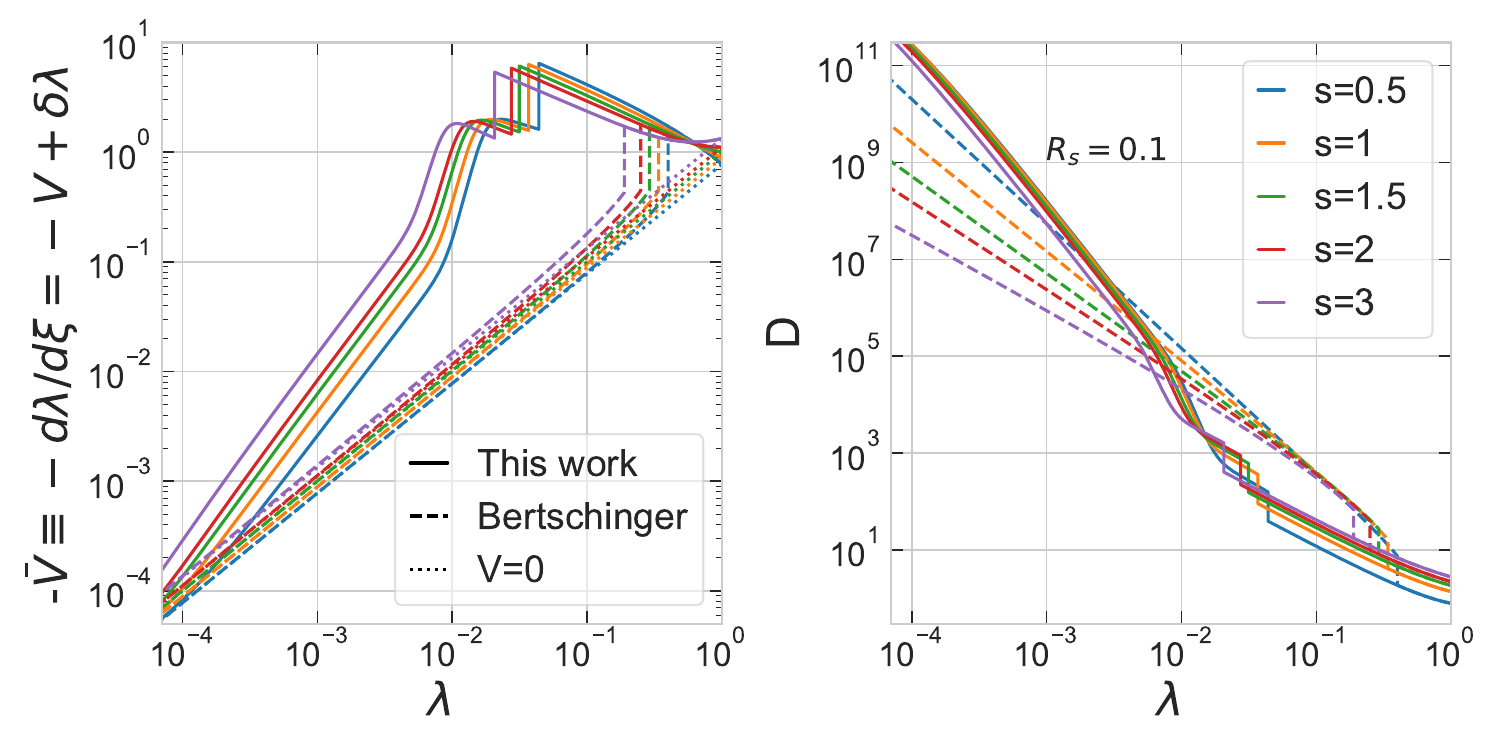}
\includegraphics[width=0.325\linewidth]{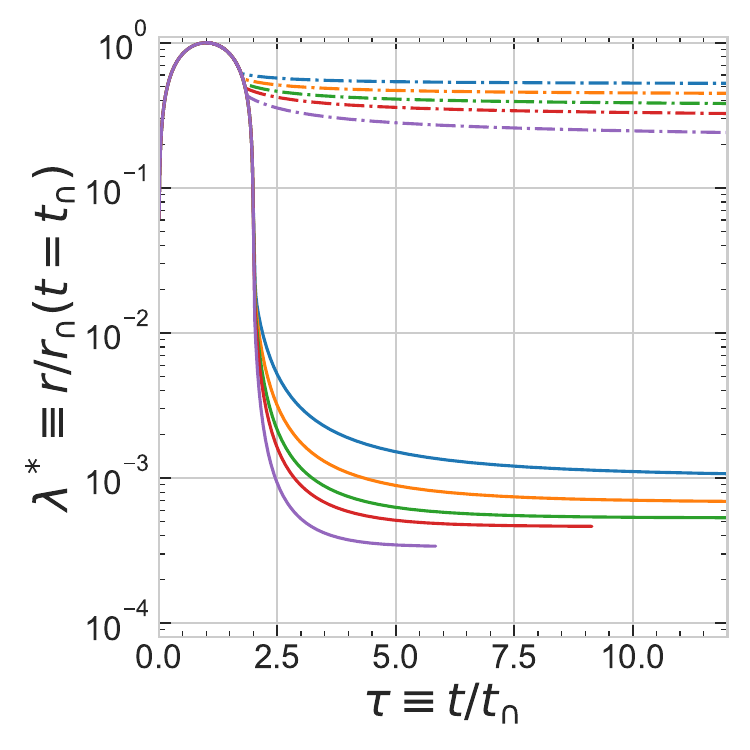}
\caption{Self-similar gas solutions are depicted, akin to those in \figref{fig:gaso-vary-s}. In a special scenario, the shock location is set at $R_{\rm s}=0.1$, however, a galaxy disk-like structure is formed at just half of this shock radius. In this scenario, we also present the self-similar solution at different accretion rates. }
\label{fig:gaso-cold-vary-s}
\end{figure}

\paragraph{Cold accretion:}
A unique scenario involves maintaining the fixed disk-like structure of the galaxy with a size equal to $5\%$ of the dark matter splashback radius and placing the shock at an extremely small radius of just twice the size of the galaxy. This arrangement allows us to mimic the cold-mode accretion of gas, since the shells retain significantly large velocity even after passing through the shock. Even the small drop in velocity at the shock is negated by the effects of cooling followed by accretion onto the pseudo-disk. As depicted in the right panel of \figref{fig:gaso-cold-vary-s}, the shells follow a nearly smooth accretion until the viscosity-like term brings them to a halt, forming a disk-like structure (see the middle panel of \figref{fig:gaso-cold-vary-s}). Even in this cold accretion scenario, the density profiles remain remarkably similar at various accretion rates, which is in contrast to the behavior observed in pressure-supported adiabatic gas. 

\begin{figure}[htbp]
\centering
\includegraphics[width=0.665\linewidth,trim={0 2cm 0 0},clip]{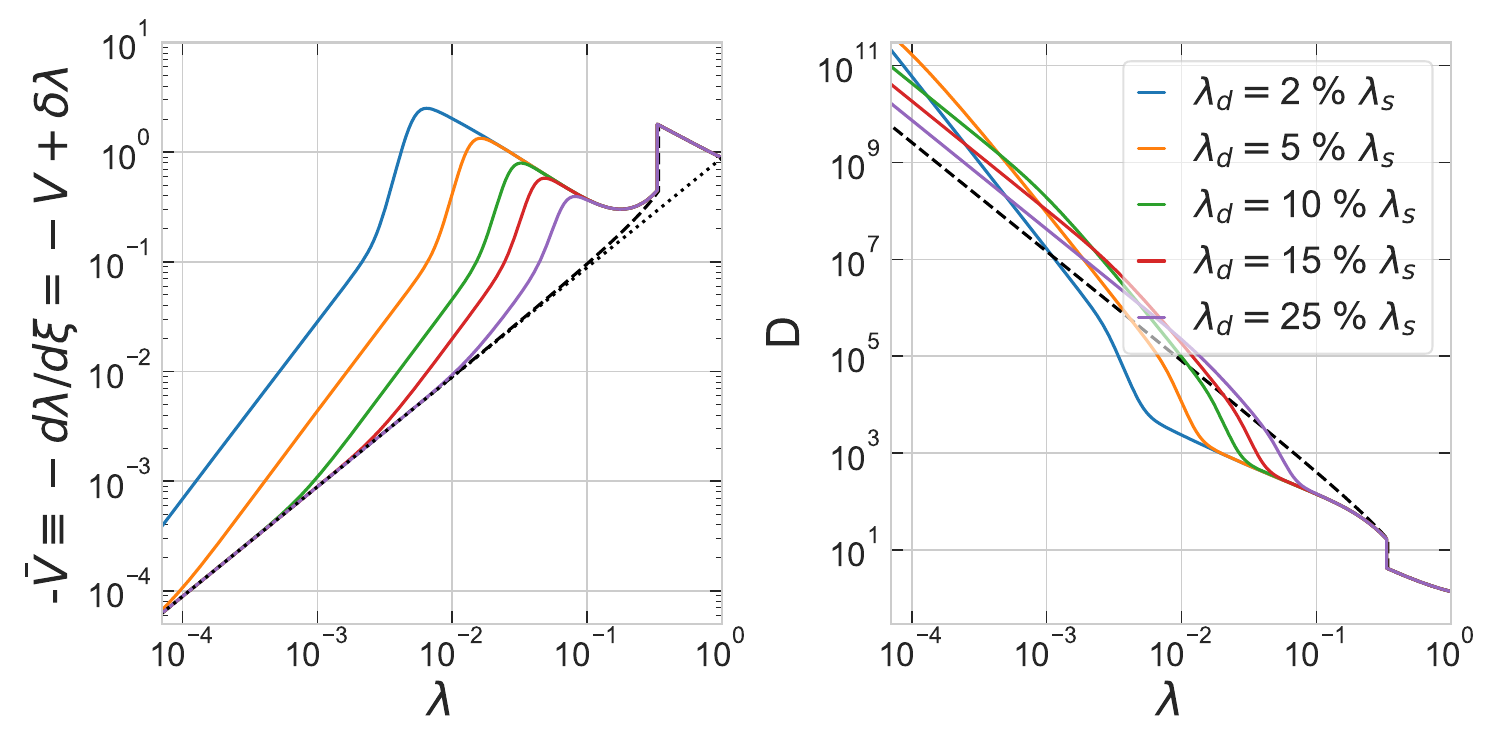} 
\includegraphics[width=0.325\linewidth,trim={0 1.8cm 0 0},clip]{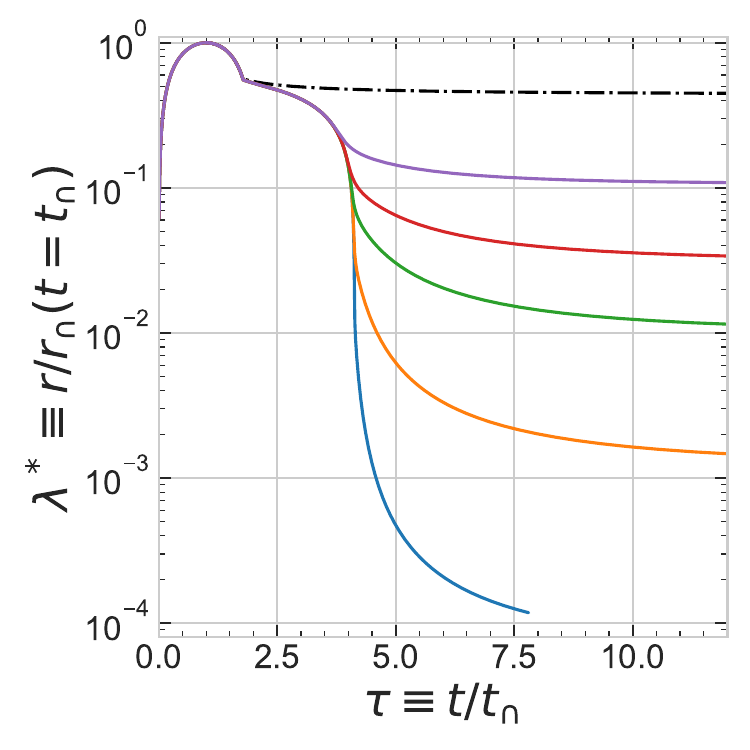}
\includegraphics[width=0.665\linewidth,trim={0 2cm 0 0},clip]{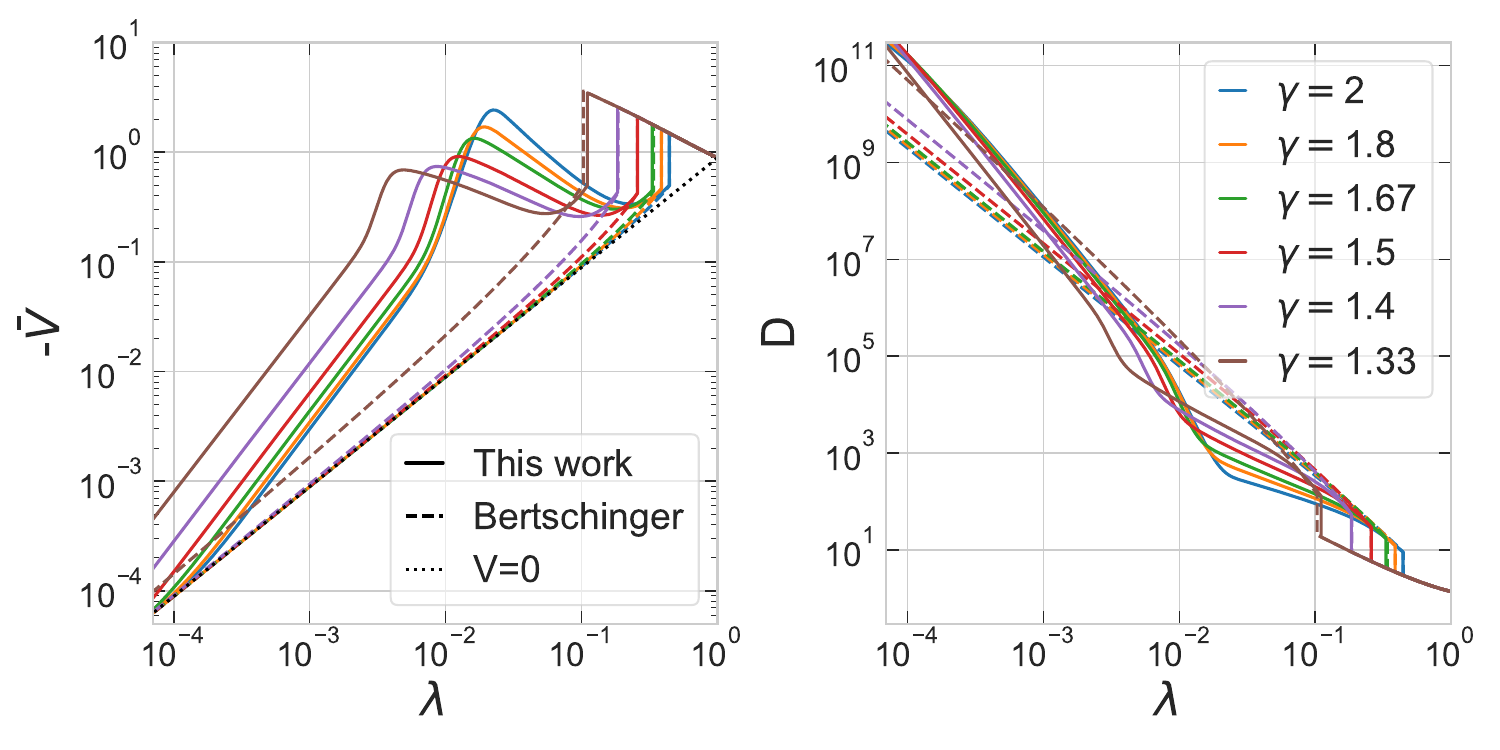}
\includegraphics[width=0.325\linewidth,trim={0 1.8cm 0 0},clip]{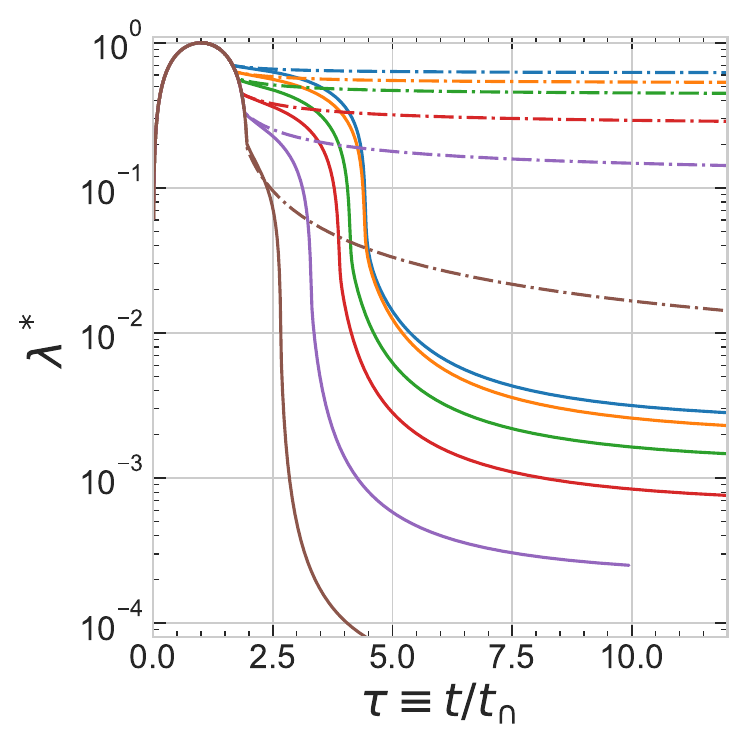}
\includegraphics[width=0.665\linewidth,trim={0 0cm 0 0},clip]{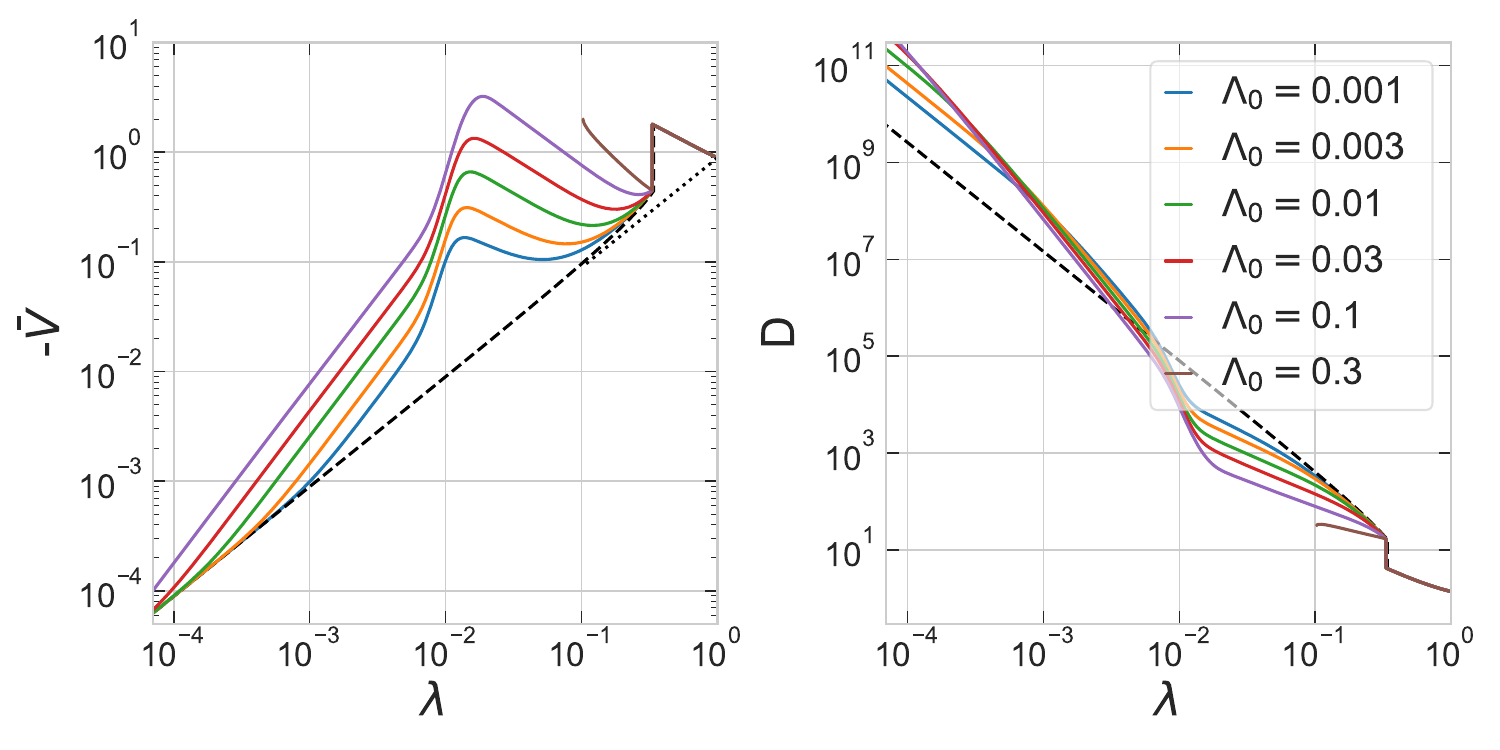}
\includegraphics[width=0.325\linewidth,trim={0 0cm 0 0},clip]{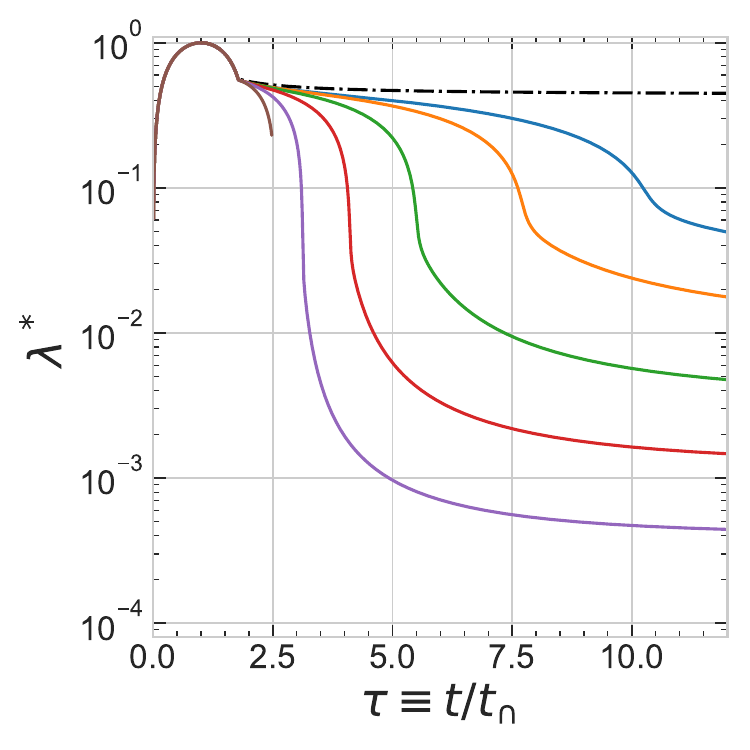}
\caption{Self-similar gas solutions are depicted, akin to those in \figref{fig:gaso-vary-s}.  In each row, one parameter is varied at a time while keeping the rest at their reference values. The top row represents variations in the size of the galaxy disk-like structure, the middle row depicts changes in the equation of state for the gas, and the bottom row shows adjustments in the cooling amplitude $\bar{\Lambda}_0$. It is important to note that while varying the gas equation of state, the shock location is also altered, as discussed in the main text.} %
\label{fig:gaso-all2}
\end{figure}

\subsubsection{Gas properties}
In this section, we demonstrate the role of various gas properties such as its equation of state, cooling behavior and the viscosity-like force it experiences in our galaxy formation model by varying only one parameter at a time.
In our model for self-similar collapse of gas, the strength of the viscosity-like term that brings the halts the shells mimicking a galaxy formation is conveniently quantified by the size of that disk-like structure $\lambda_{\rm d}$. Aligning with the usual choice in the literature \cite{2006Dekel&Birnboim}, we set this to be $5\%$ of the virial shock radius for our reference model value. By varying this parameter, we demonstrate in the top row of \figref{fig:gaso-all2}, that our galaxy formation model can be extended to produce much larger and smaller galaxies too. With the smaller value of $\lambda_{\rm d}$, the gas shells accrete quicker to more inner regions producing smaller but denser galaxies. Interestingly for larger galaxies, the profiles approach the self-similar collapse of gas with neither cooloing nor galaxy formation.

Now switching our focus to gas with different equations of state, we note that their velocities after getting shocked can vary significantly for any given choice of the shock location (see \eqn{eq:shock_jump_selfsim}. The effect of such different post-shock velocities has been studied by varying shock location and already presented in the previous section. Here we isolate the effects of the gas equation of state for shells with similar velocity inside the shock by placing the shock at a suitable radius, namely $R_{\rm s}=1.2,1.05,0.9,0.7,0.5, 0.3$ for $\gamma=2,1.8,1.67,1.5,1.4, 1.33$. These choices of shock radii are close to the fine-tuned value required for the self-similar collapse of gas with neither cooling nor galaxy disk-like structure, allowing us to compare our model with such simpler models while maintaining similar post-shock velocities.
While gases with softer equation of states are shocked at smaller radii, they also accrete onto a correspondingly smaller galaxy-like structure at nearly similar times with similar peak velocities (see middle row of \figref{fig:gaso-all2}). %

Now, we examine the impact of different cooling rates from the reference values while keeping the remaining parameters fixed, as illustrated in the bottom row of \figref{fig:gaso-all2}. In contrast to the variation in $\gamma$, for gases with different cooling rates quantified by $\bar{\Lambda}_0$, the gas undergoes shock at the same radius and accretes onto similarly sized galaxy disks in the self-similar context. However, gas that cools faster reaches much higher velocities, penetrating deeper regions of the galaxy at a much earlier time (refer to the bottom left panel of \figref{fig:gaso-all2}).
In this self-similar cooling model, we also find that when the cooling rate is set to be very high, it leads to a `cooling catastrophe' where the shells accelerate to singularity. That is the velocity shown in bottom middle panel of \figref{fig:gaso-all2} for dark brown curve corresponding to $\bar{\Lambda}_0=0.3$ accelerates to infinity at a finite $\lambda \gg \lambda_{\rm d}$; %
this prevents us from solving the evolution of gas shells further in this self-similar context to form a galaxy disk-like structure. We choose to ignore these unphysical configurations that represent the limits of our cooling model.

\subsection{Galaxy Halo co-evolution}
\label{sec:results-gas-dm-coevolve}
In this section, we 
present the self-similar solution of gas accretion onto galaxy that is co-evolving with a self-similar collapse of dark matter halo. These solutions are obtained by following the iterative procedure as described in \secref{sec:methods-interplay}. We first separately discuss the impact of the dark matter potential on the evolution of the galaxy, followed by the response of the dark matter itself to the presence of the co-evolving galaxy.

\subsubsection{Effect of background dark matter halo on gas}
\begin{figure}[htbp]
\centering
\includegraphics[width=\linewidth,trim={0 0.5cm 0cm 0},clip]{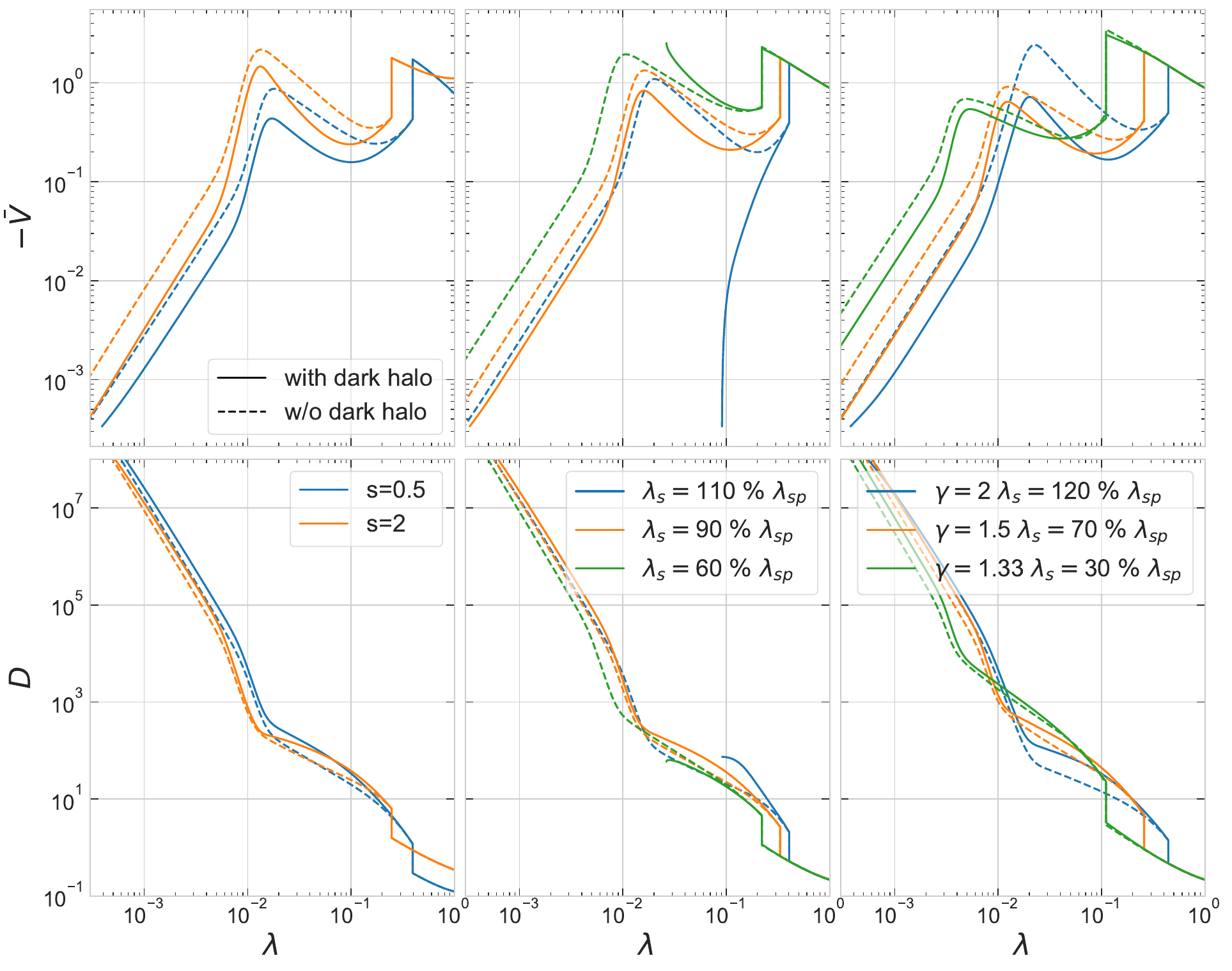}
\caption{Velocity (top row) and density (bottom row) profiles of self-similar gas in our model, co-evolved with an accreting dark matter halo are shown in solid curves. This is compared against the corresponding profiles presented in \secref{sec:results-gaso} by the dashed lines here. In the left column, this comparison is made for different accretion rates indicated by the color while keeping the rest of the model parameters to the reference values. Similarly, this response to dark matter is presented for gas that gets shocked at very different radii in the middle column, and for gas with different equations of states, getting shocked at different radii is shown in the right column. Note that the presence of a background halo, usually causes the gas shells to accrete slower, except when the shock leaves a significantly large velocity (see discussion in the main text).}
\label{fig:halo-influence-on-gas}
\end{figure}

In general, we observe that the presence of a dark matter halo tends to cause a slightly slower accretion of gas shells in most cases. This phenomenon is demonstrated by comparing the profiles of gas co-evolved with dark matter against that of gas in the absence of dark matter in the left column of \figref{fig:halo-influence-on-gas} for two distinct accretion rates. To recall, for the gas-only halo, we set the total mass within turnaround to be gas mass; on the other hand in the halo with co-evolving dark matter and gas, the cosmic baryon fraction times the total mass within turnaround is considered to be the gas mass and the rest is considered the dark matter mass. Although the velocity is marginally reduced, the overall behavior remains consistent. This is evident in the mass distribution, where a slightly larger disk is portrayed by the density profile in the bottom-left panel of \figref{fig:halo-influence-on-gas}. It is worth noting that this response is consistent across various cases presented in \secref{sec:results-gaso}, where different model parameters were considered.

An exception to this pattern arises when allowing the gas to shock at a location different from the reference choice. Specifically, when placing the shock at a larger radius while keeping the rest of the model at the reference level, the background dark matter halo causes the gas shells to potentially freeze before reaching the disk (see middle column of \figref{fig:halo-influence-on-gas}). Conversely, when placing the shock in the inner region at just $60\%$ of the dark matter splashback radius, the influence of the dark halo accelerates the gas shell accretion, triggering a `cooling catastrophe' even at the reference cooling rate.

Importantly, our findings suggest that this behavior is not solely attributed to the varying amounts of dark matter at different shock positions. Gases with different equations of state, when shocked at different radii, do not exhibit the same behavior. This is exemplified in the right column of \figref{fig:halo-influence-on-gas}, where softer gas with $\gamma=4/3$, despite being shocked at just $30\%$ of the splashback radius, experiences only a marginal reduction in velocity due to the influence of the dark halo. It is noteworthy that, despite significantly different shock radii, the post-shock velocities are similar in the top-right panel, unlike the top-middle panel. This discrepancy in post-shock velocities and, consequently, the kinetic energy of gas shells compared to the reference model, likely accounts for the unusually strong influence of dark matter.

\subsubsection{Relaxation of dark halo due to galaxy}
\begin{figure}[htbp]
\centering
\includegraphics[width=0.99\linewidth,trim={0 0cm 0cm 0},clip]{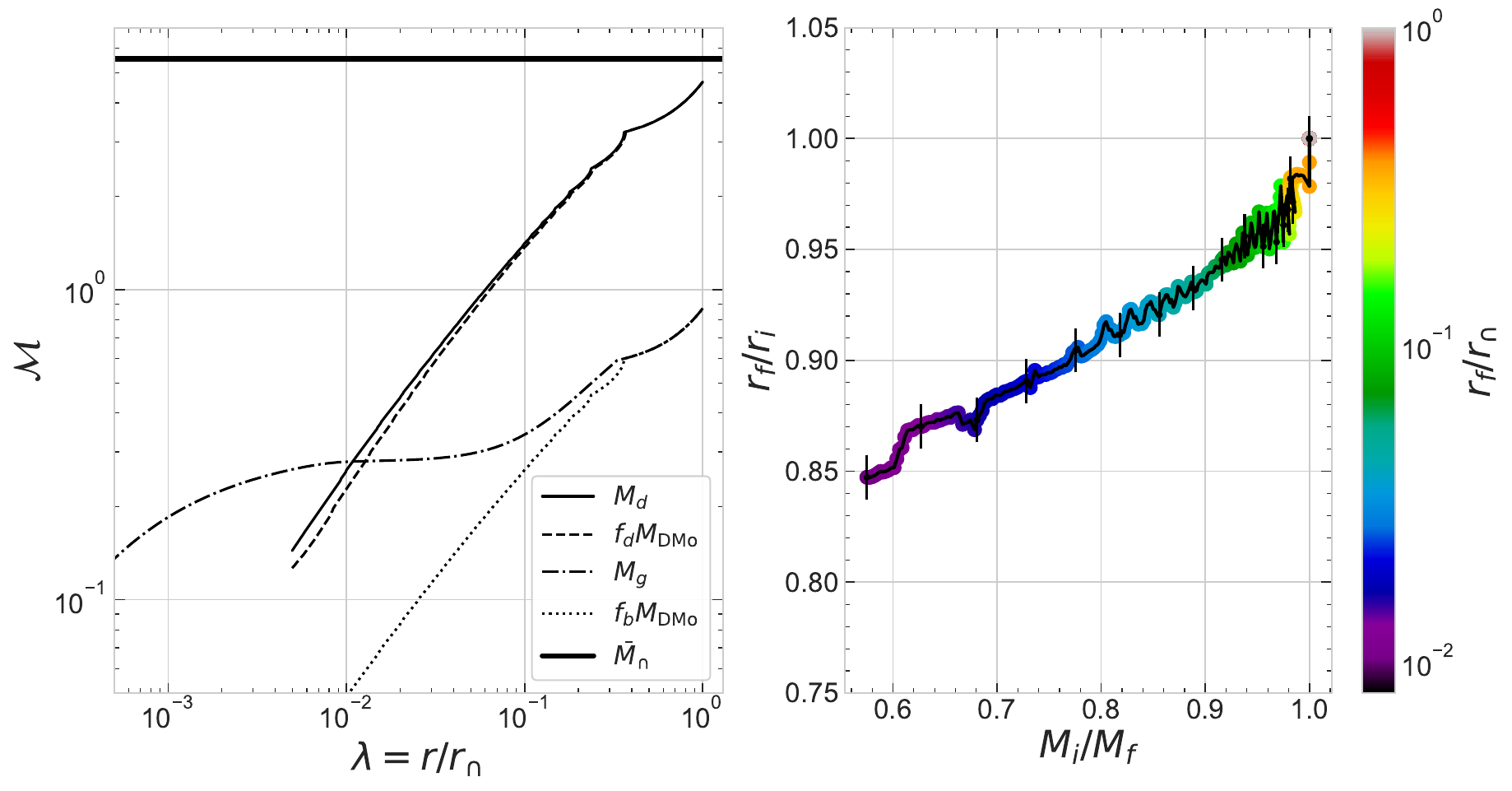}
\caption{Relaxation response of dark matter halo to galaxy formation in our self-similar model: In the left panel, a comparison of mass profiles is presented, where $M_d$ and $M_g$ denote the mass profiles of co-evolved dark matter and gas in our model. The dark matter mass profile in the absence of gas, labeled $M_{\rm DMo}$, is displayed after scaling by the cosmic dark matter fractions $f_d$ and $f_b$ for comparison. The thick horizontal line represents the total mass at the turnaround radius. The right panel illustrates the relationship between the relaxation ratio $r_f/r_i$ and the mass ratio $M_i/M_f$, as defined in \secref{sec:methods-relx-reln}. The color of the points corresponds to the radius $r_f$ scaled by the turn-around radius $r_{\smallfrown}(t)$,  and the error bars are set based on our convergence criterion. Here the dark matter mass profiles are truncated at a radius of $\lambda\sim 4 \times 10^{-3}$, since it is computationally intensive to calculate accurate profiles as $\lambda \rightarrow 0$. However in this inner region gas mass enclosed dominates over the dark matter and can be obtained with reasonable accuracy. The gas mass profile in that inner region is shown here depicting the pseudo-disk.}%
\label{fig:relx_reln}
\end{figure}

Next, we explore the relaxation response of dark matter halo to galaxy formation. As described in \secref{sec:methods-interplay}, we obtain this solution to self-similar collapse of dark matter halo co-evolving with self-similar gas accretion onto a galaxy by an iterative procedure. At each iteration, we compute the relaxation relation as described in \secref{sec:methods-relx-reln}, the relation between relaxation ratio, $r_f/r_i$ and mass ratio, $M_i/M_f$. We then continued iteration until this relaxation relation satisfies our convergence criterion, that the relative change in the relaxation ratio for $10^{-2}<r_f/r_{\smallfrown}<1$ (as shown in the right panel of \figref{fig:relx_reln}) is within $0.01$ for two consecutive iterations. Such converged solution for the mass profiles and the corresponding relaxation relation is shown in \figref{fig:relx_reln} in our model with the reference value of the parameters. By examining the mass profile ($M_d$) of dark matter that has co-evolved with a self-similar galaxy and contrasting it with the mass profile of dark matter collapse alone ($M_{\rm DMo}$) scaled by the cosmic dark matter fraction ($f_d \equiv 1-f_b$), we observe that the dark matter halo contracts in response to the galaxy's presence. 

In the context of quasi-adiabatic relaxation, as discussed in \secref{sec:methods-relx-reln}, the response of the dark matter mass profile to galaxy formation is characterized by changes in the total mass. This is attributed to the variations in the enclosed total mass within shells containing the same dark matter mass. The discrepancy in the enclosed total mass arises from the shift in baryonic mass, transitioning from the baryonic fraction of the total mass in the dark matter-only profile ($f_b \times M_{\rm DMo}$) to the gas mass profile ($M_g$) in the co-evolving system. By comparing these two curves in the left panel of \figref{fig:relx_reln}, a significant condensation of gas is evident, leading to $M_i/M_f < 1$ (refer to \secref{sec:methods-relx-reln} for detailed explanation). The corresponding relaxation relation depicted in the right panel of \figref{fig:relx_reln} supports this observation. While this relation deviates from the expectations of a standard adiabatic relaxation model, which predicts $r_f/r_i = M_i/M_f$, it exhibits a linear behavior closely resembling quasi-relaxation relations identified in full hydrodynamic simulations of cosmological volumes \cite{2023Velmani&Paranjape}. Additionally, a slight offset is noticeable, indicating that the relaxation ratio is less than unity even when the total mass ratio is unity. Similar behavior has been documented in previous full hydrodynamical simulations such as IllustrisTNG and EAGLE \cite{2023Velmani&Paranjape}. Nevertheless, it is essential to emphasize that, prior to making direct comparisons with such simulations, it is critical to integrate additional astrophysical processes, including but not limited to star formation and feedback, within this self-similar context.

\begin{figure}[htbp]
\centering
\includegraphics[height=10cm,trim={0 0 0.2cm 0},clip]{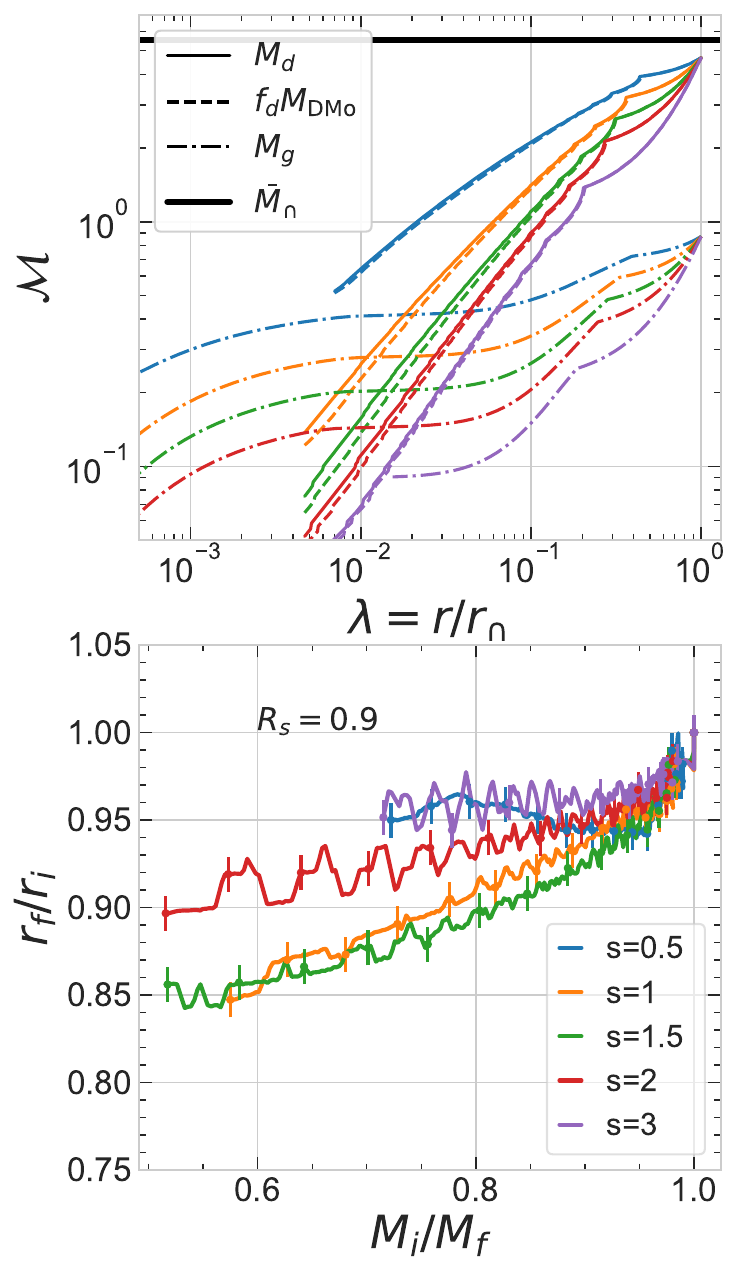}
\includegraphics[height=10cm,trim={2.2cm 0cm 0.2cm 0},clip]{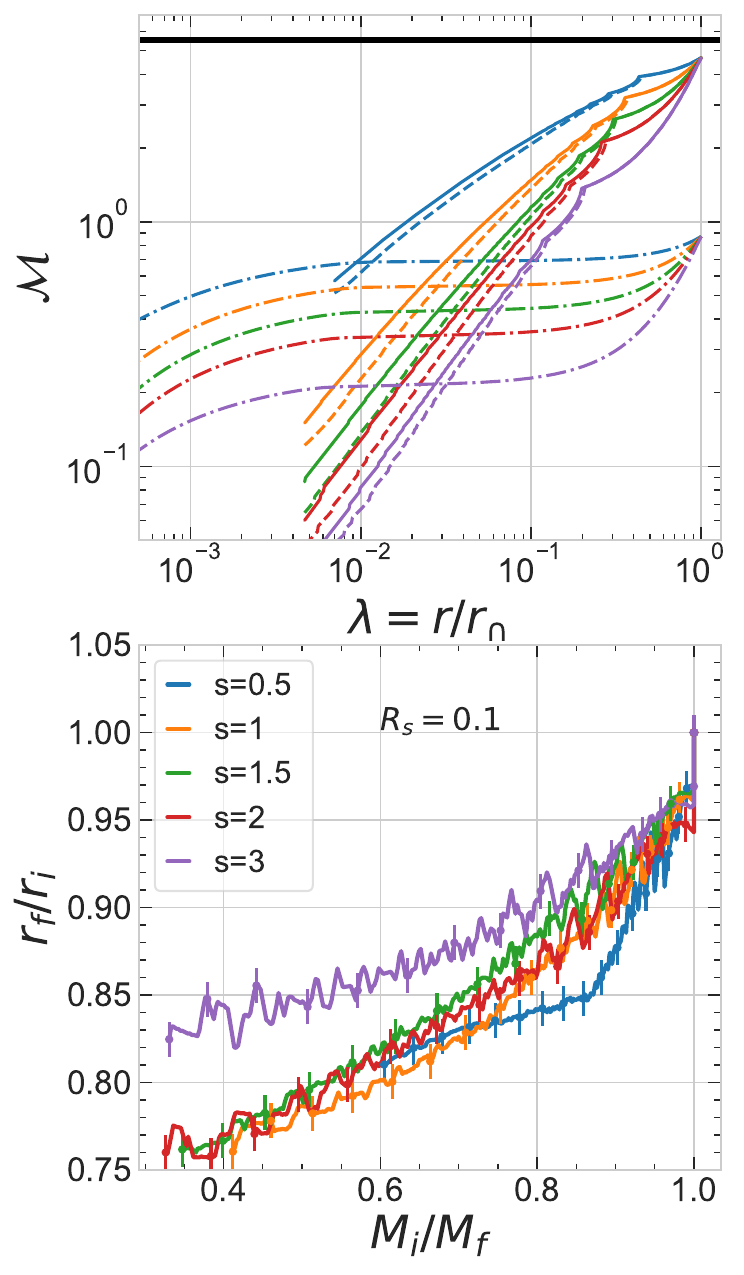}
\includegraphics[height=10cm,trim={2.2cm 0 0.2cm 0},clip]{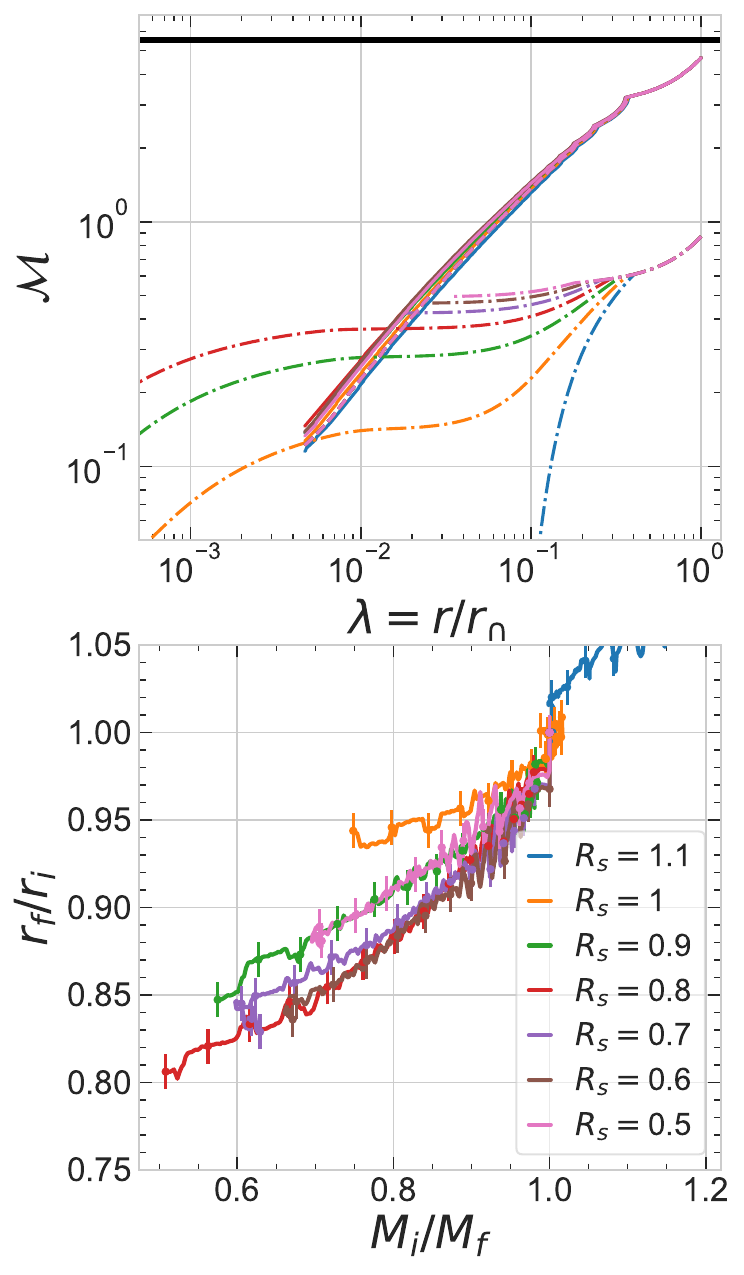}
\caption{Relaxation response of dark matter halo to galaxy formation: Investigating the impact of different parameters in our self-similar model. The upper panels feature a comparison of mass profiles, and the corresponding bottom panels depict the relationship between relaxation ratio and mass ratio, akin to \figref{fig:relx_reln}. In the left column, the accretion rate is varied while keeping the rest of the parameters fixed at reference values. In the middle column, the accretion rate is varied specifically for the case of cold mode accretion, as illustrated in \figref{fig:gaso-cold-vary-s}. In the right column, the shock location is varied as demonstrated in \figref{fig:gaso-vary-lamshsp}. }
\label{fig:relx_reln_all1}
\end{figure}

\begin{figure}[htbp]
\centering
\includegraphics[height=10cm,trim={0 0cm 0.2cm 0},clip]{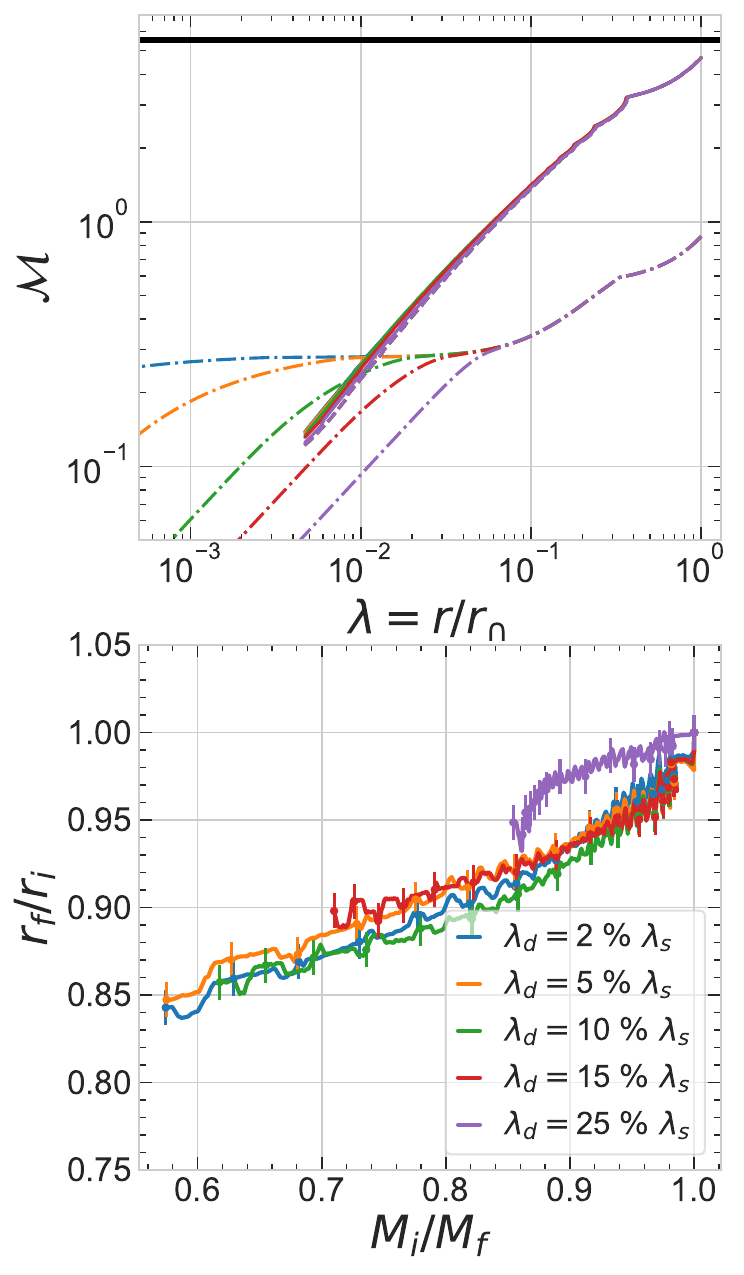}
\includegraphics[height=10cm,trim={2.2cm 0 0.2cm 0},clip]{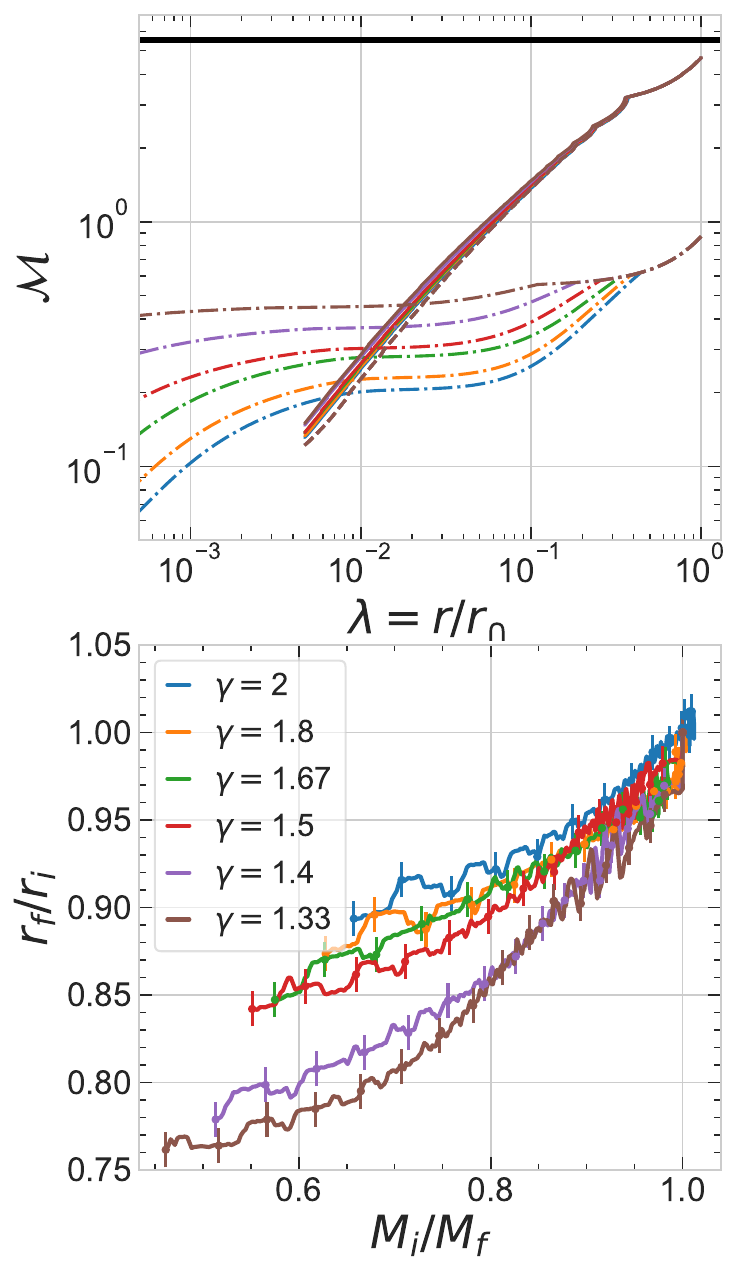}
\includegraphics[height=10cm,trim={2.2cm 0 0.2cm 0},clip]{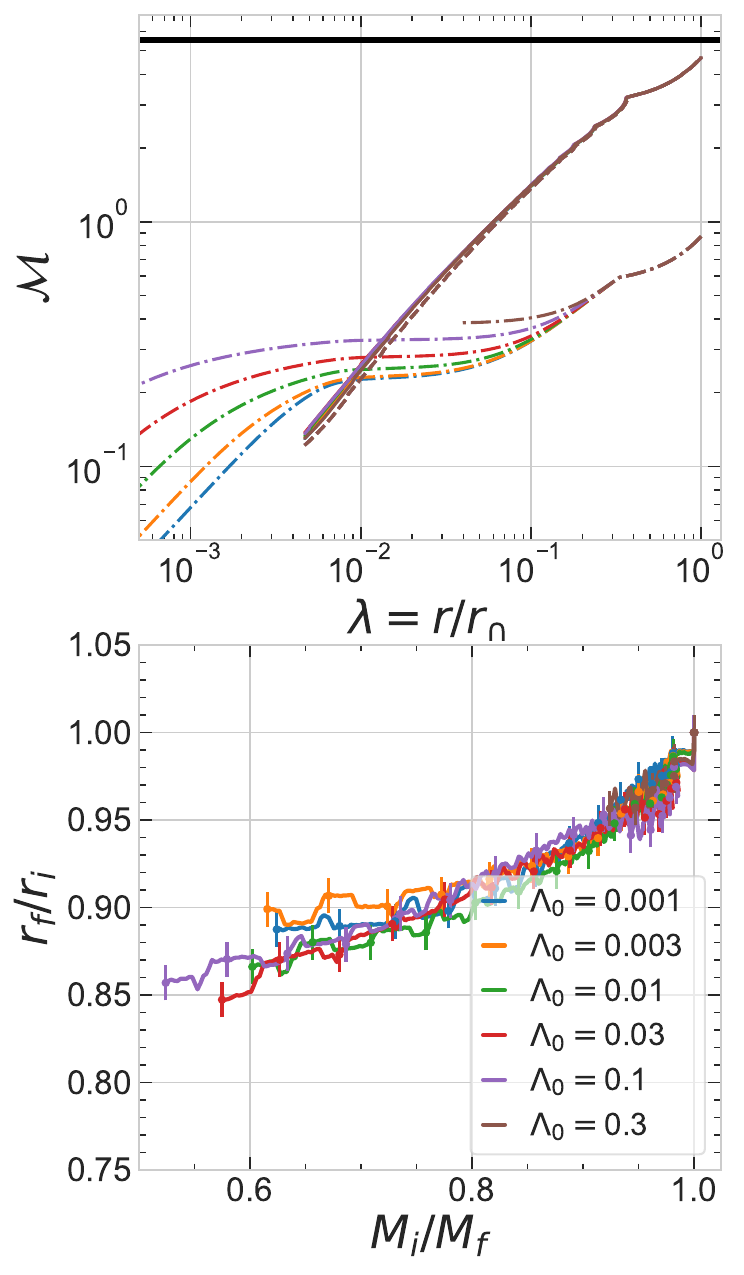}
\caption{Relaxation response of dark matter halo to galaxy formation: Investigating the impact of different parameters in our self-similar model. The upper panels feature a comparison of mass profiles, and the corresponding bottom panels depict the relationship between relaxation ratio and mass ratio, akin to \figref{fig:relx_reln}. In the left column, the size of the pseudo-disk is varied, while the middle and right columns showcase variations in the gas equation of state and cooling rate, respectively, as detailed in \figref{fig:gaso-all2}.}
\label{fig:relx_reln_all2}
\end{figure}

\paragraph{Role of halo and galaxy properties:}
By varying the parameters in our model, we have studied the universality of the relaxation behavior against such variations and the role of different halo and galaxy properties in mediating the response. This is presented in figures \ref{fig:relx_reln_all1} and \ref{fig:relx_reln_all2}, where the mass profile and the relaxation relations are presented similar to \figref{fig:relx_reln} but for cases with variation in parameters from the reference values. It is known in simulations that the gas within high-mass haloes follow shocked accretion onto the galaxy whereas the gas in low-mass haloes accrete as cold flows onto the galaxy disk. The middle column of \figref{fig:relx_reln_all1} shows the relaxation response for such low mass haloes with cold flow accretion of gas onto the galaxy. We find relaxation relation is universal with accretion rate, except when it is set to very high at $s=3$. In contrast for high mass haloes with shocked gas accretion onto galaxy, shown in the first column of that figure, the relaxation behaves very differently even at low accretion rates. Among all the parameters in our galaxy formation model, the gas equation of state parameter seems to have a larger control on the nature of the relaxation relation with a clear monotonic trend. For gas with softer equation of state, the relaxation tends to be stronger (shown in the middle panel of \figref{fig:relx_reln_all2}). 

\section{Conclusion}
\label{sec:conclusion}
We have developed a self-similar model of galaxy formation onto an isolated, evolving dark matter halo. The self-similar assumption allows us to easily track the evolution of both dark matter and gas using a system of coupled ordinary differential equations. Unlike previous work based on the self-similar assumption, a key novelty of our approach is an iterative method to simultaneously and self-consistently solve for the evolution of the dark matter and the gas. Additionally, our model is made more realistic than previous self-similar descriptions of gas infall by the inclusion of radiative cooling (through a self-similar cooling rate) and the formation of a pseudo-disk (through an effective viscosity term). With these ingredients, our model for the evolution of the gas qualitatively reproduces the results of spherical hydrodynamic simulations at a fraction of the cost. By manipulating the values of the parameters involved, our model is also able to mimic cold-mode accretion, in addition to the usual evolution of shock heated gas.

Our primary aim in constructing this model was to study the response of the dark matter to the presence of gas in the halo. This can be done using essentially the same technique as described in our previous work using cosmological hydrodynamical simulations. We found that our default configuration leads to a relaxation relation qualitatively very similar to those seen in the more realistic simulations. The flexibility of our self-similar model trivially allowed us to study the behavior of this response when varying parameters such as the accretion rate, shock location, gas adiabatic index, cooling rate amplitude and the size of the pseudo-disk. Among these, we saw that the accretion rate and adiabatic index of gas accreting in the hot mode have the most dramatic impact, while the amplitude of the cooling rate has relatively little impact.

Our model provides a framework to systematically explore the coupled impact of multiple astrophysical processes on the mass and velocity profile of dark matter in galactic halos, some of which we studied above. 
Additional applications include studying the impact of baryons on rotation curves and gravitational lensing signals.
Our model also has potential applications in `baryonification' exercises while building emulators for cosmological parameter inference.
In order to achieve this with sufficient realism, however, we must also incorporate the effects of star formation and the development of a central black hole in the galactic system, along with the associated feedback of thermal and/or kinetic energy. We leave the development of these additional effects to future work.

This self-similar approach potentially provides certain advantages over 1D spherical hydrodynamical simulations. Firstly, the sole computationally intensive task in this self-similar model lies in obtaining the dark matter profile deep within the halo. This might provide a computational advantage over 1D simulations, especially when exploring a large space of models. More importantly, as a semi-analytic approach, this self-similar model has the potential to yield straightforward analytical relations. These derived relations serve as valuable tools in developing quantitative models for relaxation. For instance, they facilitate the quantification of relaxation as a function of halo/galaxy properties, providing a means to extract insightful and applicable insights on relaxation.

However, there are some caveats in this self-similar approach.
Foremost is the inability of this self-similar model to precisely replicate the NFW density profile in the virial region, since it relies on a power-law-like mass growth history, and adheres to EdS-cosmology rather than $\Lambda$CDM.
Also, major merger events induce shocks that can significantly alter the properties of the accretion shock \cite{2020Zhang_etal_merger_shocks,2020ShiNagai_etal_merger_shocks}, potentially challenging these predictions from self-similar spherical collapse models.

It may be possible to incorporate the key missing physics into the current framework given its flexibility. Cold accretion may be modelled by another species with a DM-like equation of state but subject to viscosity. A more realistic dark matter profile in the Virial regions might be possible by adding viscosity to dark matter shells. Shi (2023) \cite{2023Shi_iter_approach} has used an iterative mean-field approach to generalize spherical collapse to $\Lambda$CDM cosmology and realistic mass accretion histories. We leave the exploration of these ideas to future work.

\section*{Data availability}
No new data were generated during the course of this research.

\acknowledgments

The research of AP is supported by the Associates scheme of ICTP, Trieste.

\bibliography{references}

\end{document}